\documentclass[aps,11pt,prd,nofootinbib,showpacs,superscriptaddress,preprint]{revtex4-1}
\usepackage{booktabs,amsmath}
\usepackage{amssymb}
\usepackage{bm}
\usepackage{amsmath}
\usepackage{array}
\usepackage{mathtools}
\usepackage{amsthm}
\usepackage{newlfont}
\usepackage[utf8]{inputenc}
\usepackage{esdiff}
\usepackage{multirow}
\usepackage{graphics}
\usepackage{graphicx}
\usepackage{ae,aecompl,color}

\usepackage[bookmarks=true,bookmarksopen=true,bookmarksnumbered=true,bookmarksopenlevel=3]{hyperref}

\makeatletter
\DeclareRobustCommand*{\bfseries}{%
  \not@math@alphabet\bfseries\mathbf
  \fontseries\bfdefault\selectfont
  \boldmath
}
\makeatother

\begin{document}
\newcommand{\bd}{\begin{document}}
\newcommand{\ed}{\end{document}}
\newcommand{\bc}{\begin{center}}
\newcommand{\ec}{\end{center}}
\newcommand{\bfr}{\begin{flushright}}
\newcommand{\efr}{\end{flushright}}
\newcommand{\lt}{\left}
\newcommand{\rt}{\right}
\newcommand{\vs}{\vspace}
\newcommand{\hs}{\hspace}
\newcommand{\beq}{\begin{equation}}
\newcommand{\eeq}{\end{equation}}
\newcommand{\lb}{\linebreak}
\newcommand{\pb}{\pagebreak}
\newcommand{\mb}{\makebox}
\newcommand{\fb}{\framebox}
\newcommand{\mc}{\multicolumn}
\newcommand{\ben}{\begin{enumerate}}
\newcommand{\een}{\end{enumerate}}
\newcommand{\bit}{\begin{itemize}}
\newcommand{\eit}{\end{itemize}}
\newcommand{\un}{\underline}
\newcommand{\lefq}{\lefteqn}
\newcommand{\ba}{\begin{array}}
\newcommand{\ea}{\end{array}}
\newcommand{\beqa}{\begin{eqnarray}}
\newcommand{\eeqa}{\end{eqnarray}}
\newcommand{\beqas}{\begin{eqnarray*}}
\newcommand{\eeqas}{\end{eqnarray*}}
\newcommand{\bfg}{\begin{figure}}
\newcommand{\efg}{\end{figure}}
\newcommand{\bds}{\begin{displaymath}}
\newcommand{\eds}{\end{displaymath}}
\newcommand{\btb}{\begin{tabbing}}
\newcommand{\etb}{\end{tabbing}}
\newcommand{\para}{\parallel}
\newcommand{\pad}{\partial}
\newcommand{\nn}{\nonumber}
\newcommand{\la}{\leftarrow}
\newcommand{\ra}{\rightarrow}
\newcommand{\lgla}{\longleftarrow}
\newcommand{\lgra}{\longrightarrow}
\newcommand{\La}{\Leftarrow}\newcommand{\Ra}{\Rightarrow}
\newcommand{\Lra}{\Leftrightarrow}
\newcommand{\Lgla}{\Longleftarrow}
\newcommand{\Lgra}{\Longrightarrow}
\newcommand{\lan}{\langle}
\newcommand{\ran}{\rangle}
\renewcommand{\a}{\alpha}
\renewcommand{\b}{\beta}
\newcommand{\g}{\gamma}
\newcommand{\G}{\Gamma}
\renewcommand{\d}{\delta}
\newcommand{\eps}{\epsilon}
\newcommand{\Th}{\Theta}
\newcommand{\s}{\sigma}
\newcommand{\lam}{\lambda}
\newcommand{\D}{\Delta}
\newcommand{\vare}{\varepsilon}
\newcommand{\pr}{\prime}
\newcommand{\ro}{\rho}
\newcommand{\nab}{\nabla}
\newcommand{\m}{\mu}
\newcommand{\n}{\nu}
\newcommand{\Sg}{\Sigma}
\newcommand{\p}{\pi}
\newcommand{\R}{I\!\!R}
\newcommand{\om}{\omega}
\newcommand{\Om}{\Omega}
\newcommand{\ze}{\zeta}
\newcommand{\vart}{\vartheta}
\newcommand{\tri}{\triangle}
\newcommand{\f}{\frac}
\newcommand{\iny}{\infty}
\newcommand{\pro}{\propto}
\renewcommand{\arraystretch}{1.25}
\title{Quantum phase transitions in the noncommutative Dirac Oscillator} 
%
%
\author{\textsc{O.~Panella}}
\affiliation{Istituto Nazionale di Fisica Nucleare, Sezione di Perugia, Via A.~Pascoli, I-06123 Perugia, Italy}
\email[({\bf Corresponding Author}) Email: ]{orlando.panella@pg.infn.it }

\author{\textsc{P.~Roy}}
\affiliation{Physics and Applied Mathematics Unit, Indian Statistical Institute, Kolkata, India}

\date{\today}

\begin{abstract}
We study the (2+1) dimensional Dirac oscillator in a homogeneous  magnetic field in the non-commutative plane. It is shown that the effect of non-commutativity is twofold: $i$)  momentum non commuting coordinates simply shift  the critical value ($B_{\text{cr}}$) of the magnetic field at which  the well known left-right chiral quantum phase transition takes place (in the commuting phase); $ii$) non-commutativity in the space coordinates induces a new critical value of the magnetic field, $B_{\text{cr}}^*$, where there is a second quantum phase transition (right-left), --this critical point disappears in the commutative limit--. The change in chirality associated with the magnitude of the magnetic field is examined in detail for both critical points. The phase transitions are described in terms of the magnetisation of the system.  Possible applications to the physics of silicene and graphene are briefly discussed.
\end{abstract}

\maketitle

\section{Introduction} Recent studies in string theory and quantum gravity  indicate that space-time can be non-commutative \cite{Connes:1998aa,Seiberg:1999aa,Douglas:2001aa,Szabo:2003aa}. Subsequently quantum mechanics in such a space-time has been studied by a number of authors to determine the role of noncommutativity parameters on a variety of physical  observables \cite{Duval:2000aa,Gamboa:2001aa,Gamboa:2001ab,Nair:2001aa,Gamboa:2002aa,Smailagic:2002aa,Smailagic:2002ab,Bellucci:2001aa,Bertolami:2005aa,Horvathy:2010wv}. 

The Dirac oscillator on the other hand is one of the very few relativistic systems which is exactly solvable \cite{Ito:1967aa,Cook:1971aa,Moshinsky:1989aa,Benitez:1990aa,Rozmej:1999aa,Mirza:2004aa,Sadurni:2010aa,Boumali:2013aa}.  In mathematical physics the Dirac oscillator has become a paradigm for the realization of covariant quantum models and it has found applications both in nuclear~\cite{Faessler:2005aa,Grineviciute:2012aa,Munarriz:2012aa} and subnuclear~\cite{Romera:2011aa,Wang:2012aa} physics as well as in quantum optics~\cite{Bermudez:2007ab,Dodonov:2002aa,Lamata:2007aa}.  Very recently a one dimensional version of the Dirac oscillator has been realized experimentally~\cite{Franco-Villafane:2013aa} for the first time. Using  a microwave setup the Dirac Oscillator has been modeled by a chain of coupled disks with high index of refraction and only nearest neighbor interactions sandwiched between metallic plates. By adjusting properly the coupling between the disks  this system reproduces the spectrum of the one dimensional Dirac oscillator. In \cite{Franco-Villafane:2013aa} the authors also discuss prospects to realize in the near future the two dimensional version of the Dirac oscillator which may be feasible using networks of microwave coaxial cables~\cite{Hul:2004aa,Sadurni:2010fk,Hul:2005uq}. It is interesting to note that a further interaction in the form of a homogeneous magnetic field can still be incorporated in the Dirac oscillator keeping the system still exactly solvable \cite{Bermudez:2007aa,Bermudez:2007ab,Bermudez:2008aa,Mandal:2010aa,Mandal:2012aa} and this combined system has quite interesting properties. In some recent papers it has been shown that for this combined system there is a chirality phase transition if the magnitude of the magnetic field either exceeds or is less than a critical value $B_{\text{cr}}$ (which also depends on the oscillator strength) \cite{Bermudez:2008ab,Quimbay:2013aa}. A consequence of this chirality phase transition is that the spectrum is different for $B>B_{\text{cr}}$ and $B<B_{\text{cr}}$, $B$ being the magnetic field strength. 

Our objective here is to analyse the same system,  a 2D Dirac oscillator within a constant magnetic field, but in the framework of non-commutative space and momentum coordinates. It will be shown that in this case the critical value of the magnetic field depends not only on the oscillator strength but on the non-commutativity parameters as well. Another interesting consequence of the non-commutative scenario which we point out is that, apart from the two left- and right-chiral phases of the  commutative case, there is also a \emph{new} third left phase and so a second quantum phase transition (right-left) which both cease to exist as the space non-commutativity parameter vanishes. We shall discuss the spectrum and the degeneracy of the various energy levels in all three phases. Another point which we address is the following: how to characterise the chirality phase transition? It will be shown that the chirality phase transition can be described in terms of the magnetization of the system.  We shall also examine the quantum phase transition(s) in the context of new 2-dimensional materials, namely, graphene~\cite{Geim:2007aa,Castro-Neto:2009aa,Semenoff:1984aa} and silicene~\cite{Lalmi:2010kx,Vogt:2012vn,Fleurence:2012ys,Lin:2012zr}. Charge carriers in Graphene, a 2-dimensional sheet of carbon atoms, are  known to be described at the $K$, and $K'$ points of the Brillouin zone by an effective massless Dirac equation. Silicene, the silicon counterpart of graphene, is known to be described at the $K,K'$ points by a massive Dirac equation. In consideration of the fact that in \cite{Quimbay:2013ab} it has been conjectured that in graphene a Dirac oscillator coupling may arise as a consequence of an effective internal magnetic field due to the motion of the electrons in the planar hexagonal lattice of the carbon atoms, it is clear that the exact solutions presented in this work may have direct relevance to experimentally accessible physical systems (like silicene) of great interest to the scientific community, independently of the existence or not of the (admittedly speculative) non commutative scenario. Our solutions being derived for a massive Dirac oscillator would have to be considered in the massless limit to be applicable to graphene.   Although we have studied the system in a non-commutative space, our analysis can indeed be applied to the commutative case as well by letting the non-commutativity parameters vanish. 

The organisation of the paper is as follows: in section \ref{formu} we have presented outline of the problem; in sections \ref{weak}, \ref{inter} and \ref{strong} we discuss the spectrum in three phases, namely, the weak, the intermediate and the strong magnetic field; in section \ref{discussion} we discuss how to characterise the chirality phase transition as the strength of the magnetic field varies along with possible applications to silicene and graphene, and finally section \ref{conclusions} is devoted to our conclusions.

\section{Formulation of the problem}\label{formu}
To begin with we note that the Hamiltonian for the $(2+1)$ dimensional Dirac oscillator in the noncommutative plane in the presence of a homogeneous magnetic field is given by
\beq
H=c\bm{\sigma}.(\bm{\hat p}-im\omega\beta \bm{\hat x}+\f{e}{c}\bm{{\hat A}})+\b mc^2\label{h1}\,,
\eeq
where $c$ is the velocity of light and $\bm{\sigma}$, $\beta=\sigma_z$ denote Pauli matrices. We now choose the vector potential as
\beq
\bm{\hat{A}}=(-B{\hat y}/2,B{\hat x}/2,0)\, .
\eeq
The commutation relation between coordinates and momenta are given by
\beq
\label{NCEQ}
[{\hat x},{\hat y}]=i\theta,~~~~[{\hat p}_x,{\hat p}_y]=i\eta,~~~~[{\hat x}_i,{\hat p}_j]=i\hbar(1+\f{\theta\eta}{4\hbar^2})\delta_{ij}, \qquad \theta,\eta \in \mathbb{R}\, .
\eeq
Then written in full the above Hamiltonian reads
\beq
H= c\left(\ba{cc} mc & {\hat\Pi}_-\\ {\hat\Pi}_+ & -mc\ea\right)\label{h2},
\eeq
where $\Pi_{\pm}$ are given by
\beq
\ba{lcl}
{\hat\Pi}_-=\displaystyle {\hat p}_x-i{\hat p}_y-i\left(\f{eB}{2c}-m\omega\right)\,{\hat x}-\left(\f{eB}{2c}-m\omega\right)\,{\hat y}\,;\\
{\hat\Pi}_+=\displaystyle {\hat p}_x+i{\hat p}_y+i\,\left(\f{eB}{2c}-m\omega\right)\,{\hat x}-\left(\f{eB}{2c}-m\omega\right)\,{\hat y}\,.
\ea
\eeq
It is now necessary to express the noncommuting coordinates and momenta in terms of commuting ones. This can be achieved using the Seiberg-Witten map and are given by
\beq
\ba{lcl}
{\hat x}=\displaystyle {x-\f{\theta}{2\hbar}p_y,~~~~{\hat p}_x=p_x+\f{\eta}{2\hbar}y}\,,\\
{\hat y}=\displaystyle{y+\f{\theta}{2\hbar}p_x,~~~~{\hat p}_y=p_y-\f{\eta}{2\hbar}x}\label{rel}\,,
\ea
\eeq
where $(x,y)$ and $(p_x,p_y)$ denote commuting coordinates and momenta. Now using the relation (\ref{rel}) the Hamiltonian (\ref{h2}) can be written as
\beq
H=c\left(\ba{cc} mc & {\hat\Pi}_-\\ {\hat\Pi}_+ & -mc\ea\right)\label{h3}\,,
\eeq
where ${\hat\Pi}_{\pm}$ are more conveniently written after introducing the following frequencies out of the parameters of the problem: 
\begin{equation}
\tilde{\omega}=\frac{eB}{2mc},\, \qquad\qquad \omega_\theta = \frac{2\,\hbar}{m\theta},\,\qquad \qquad \omega_\eta=\frac{\eta}{2\hbar m}.
\end{equation}
Then we find:
\beq
\label{pi}
{\hat\Pi}_{\pm} = \left (1- \displaystyle\frac{\tilde{\omega}-\omega}{\omega_\theta}\right)(p_x \pm i\, p_y)\, \pm\, i\, m\, (\tilde{\omega} -\omega -\omega_\eta) (x \pm i y)\,.
\eeq

From the above relations one finds that there is a critical value of the magnetic field $(B_{\text{cr}}$)
\beq
\label{bc}
B_{\text{cr}}=\frac{2 mc}{e} (\omega +\omega_\eta)=\f{2c}{e}(m\omega+\f{2\eta}{\hbar}) \,,
\eeq
such that for $B=B_{\text{cr}}$ (or $\tilde{\omega} =\omega +\omega_\eta$) there are no interactions in the model and the Hamiltonian represents a free particle (only kinetic energy). Note that the critical magnetic field $B=B_{\text{cr}}$ actually depends on the oscillator frequency as well as the momentum non-commutativity parameter ($\eta$ or $\omega_\eta$). We see therefore that the momentum non-commutativity  parameter ($\eta$) shifts the value of the critical filed relative to the value in the commutative case, where the system is well known to undergo a quantum phase transition.
  
More importantly we note that space non commutativity ($\theta \ne 0$) introduces an additional critical value for the magnetic field ($B_{\text{cr}}^*$) which corresponds to the the condition $\tilde{\omega}=\omega+\omega_\theta$:
\beq
\label{bcrit*}
B_{\text{cr}}^*=\frac{2 mc}{e} (\omega +\omega_\theta)=\f{2c}{e}(m\omega+\f{\hbar}{\theta}) \, .
\eeq
 At this value of the magnetic field the kinetic part of the Hamiltonian will vanish. Physically we can visualise the system at this critical value as being made up by \emph{only potential energy}. This second critical point signals a new quantum phase transition which is \emph{absent} in the commutative limit, whereas the quantum phase transition at $B=B_{\text{cr}}$ is only shifted by the momentum non commutativity parameter $\eta$ (and this shift goes smoothly to zero when $\eta \to 0$). 
 
In the following section we shall analyse the spectrum as a function of the magnetic field separately in each of the three regions: when $B<B_{\text{cr}}$, $B_{\text{cr}} < B < B_{\text{cr}}^*$ and $ B> B_{\text{cr}}^*$. 

We shall discuss the problem according to the magnitude of the magnetic field (or the cyclotron frequency $\tilde{\omega}$) relative to the other parameters (Dirac oscillator frequency $\omega$ and NC frequencies $\omega_\eta,\omega_\theta$). In this regard we note that as the non commutative (NC) parameters $\eta,\theta \to 0$ (the commutative limit) the frequency $\omega_\eta$ vanishes, while $\omega_\theta \to \infty$.
Therefore without any loss of generality we may assume the following relation between 
$\omega_\eta$, and $\omega_\theta$:
\begin{equation}
\omega_\eta <  \omega_\theta\, .
\end{equation}

\section{Weak Magnetic Field: $B<B_{\text{cr}}$ ($\tilde{\omega} < \omega +\omega_\eta$)}
\label{weak}

In this case we find from Eq.~\eqref{pi}: 
\beq
\label{pi1}
{\hat\Pi}_\pm=\lambda(p_x\pm i\,p_y)\mp i\nu (x \pm y)\, ,
\end{equation}
\begin{equation}
\lambda= 1- \displaystyle\frac{\tilde{\omega}-\omega}{\omega_\theta} > 0\, , \qquad \qquad \nu= m (\omega +\omega_\eta -\tilde{\omega}) > 0\,.
\eeq
To this end let us define the following set of creation and annihilation operators: 
\beq\label{aadag1}
\ba{lcl}
\displaystyle{a^{\phantom{\dag}}_x=\f{i}{\sqrt{2\nu\lambda\hbar}}(\lambda p_x-i\nu x),~~~~a_x^\dag=\f{-i}{\sqrt{2\lambda\nu\hbar}}(\lambda p_x+i\nu x})\,,\\
\displaystyle{a^{\phantom{\dag}}_y=\f{i}{\sqrt{2\nu \lambda \hbar}}(\lambda p_y-i\nu y),~~~~a_y^\dag=\f{-i}{\sqrt{2\lambda\nu\hbar}}(\lambda p_y+i\nu y)}\,,
\ea
\eeq
and then the so called circular (or chiral) annihilation and creation operators:
\beq\label{alr}
\ba{lcl}
a^{\phantom{\dag}}_L=\displaystyle{\f{1}{\sqrt{2}}(a_x+ia_y),~~~~a_L^\dag=\f{1}{\sqrt{2}}(a_x^\dag-ia_y^\dag)}\,,\\
a^{\phantom{\dag}}_R=\displaystyle{\f{1}{\sqrt{2}}(a_x-ia_y),~~~~a_R^\dag=\f{1}{\sqrt{2}}(a_x^\dag+ia_y^\dag)}\,.
\ea
\eeq
It can easily be verified that the operators in (\ref{alr}) satisfy the following commutation relations 
\beq
[a^{\phantom{\dag}}_R,a_R^\dag]=[a^{\phantom{\dag}}_L,a_L^\dag]=1,~~~~[a^{\phantom{\dag}}_R,a^{\phantom{\dag}}_L]=[a^{\phantom{\dag}}_R,a_L^\dag]=[a_R^\dag,a^{\phantom{\dag}}_L]=[a_R^\dag,a_L^\dag]=0\,.
\eeq
Thus $(a_{R,L}^\dag,a^{\phantom{\dag}}_{R,L})$ represent creation and annihilation operators for a pair of independent harmonic oscillators. We shall now analyse the spectrum using the creation and annihilation operators defined in (Eq.~\ref{alr}).

As a consequence we obtain:
\beq
\label{LI}
{\hat\Pi}_+=-2\,i\,\sqrt{\nu\lambda\hbar}\, a_L^{\phantom\dag},\qquad \qquad{\hat\Pi}_-=+2\, i\, \sqrt{\nu\lambda\hbar}\,a_L^\dag\, .
\eeq
The eigenvalue equation can be written as
\beq
+2\, i\, \sqrt{\nu\lambda\hbar}\, a_L^\dag \, \psi_L^{(2)}=\eps_-\psi_L^{(1)},~~~~ -2\, i\, \sqrt{\nu\lambda\hbar}\, a_L^{\phantom{\dag}}\, \psi_L^{(1)}=\eps_+\psi_L^{(2)},~~~~\eps_\pm=\f{E\pm mc^2}{c}\label{eigen}\,.
\eeq
Let us now examine whether there are solutions   with energies $E=\pm mc^2$ i.e, $\eps_{\pm}=0$. 
It turns out to be convenient to work in standard polar coordinates:
\[
r=\sqrt{x^2+y^2}\,, \qquad \qquad  \varphi  = \arctan\left(\frac{y}{x}\right)\,,  \qquad   \to \qquad (x,y) = (r\cos\varphi,r \sin\varphi)\,,
\]
and then we can easily find the form of the annihilation and creation operators:
\begin{eqnarray}
a_L^{\phantom{\dag}} \,& =&\, \frac{e^{i\varphi}}{2\sqrt{\nu\lambda \hbar}} \,\left\{ \phantom{+} \lambda \hbar \, \left( \partial_r +\frac{i}{r}\, \partial_{\varphi} \right) +\nu r \right\}\,, \nonumber \\
a_L^{{\dag}} \, &=&\, \frac{e^{-i\varphi}}{2\sqrt{\nu\lambda \hbar}} \,\left\{ - \lambda \hbar\,\left( \partial_r -\frac{i}{r} \, \partial_{\varphi} \right) +\nu r \right\}\, , \nonumber
\end{eqnarray}  


\beq
\label{zeroLmodes}
\ba{lcl}
a_L^{\phantom{\dag}}\,\psi_L^{(1)}=0,\quad \to\quad \left \{ 
 \partial_r +\frac{i}{r}\, \partial_{\varphi} +\frac{\nu}{\lambda \hbar} r 
\right \}\psi_L^{(1)}(r,\varphi)=0 \, ,\\
a_L^{{\dag}}\,\psi_L^{(2)}=0,\quad \to\quad \left\{      
  \partial_r -\frac{i}{r}\, \partial_{\varphi} -\frac{\nu}{\lambda \hbar} r 
\right\}\psi_L^{(2)}(r,\varphi)=0 \, .
\ea
\eeq

Let us first consider the first of Eqs.~\eqref{zeroLmodes}. This is a first order partial differential equation which can be easily solved by the method of separation of the variables. For the wave function  we  make the  ansatz  $\psi_L^{(2)} = u(r) Y(\varphi) $ and find with straightforward calculations that $u(r) = r^M e^{-\frac{\nu}{2\lambda \hbar} r^2} +c_1$ and $Y(\varphi) = e^{iM\varphi}+c_2$ with $M,c_1$ and $c_2$ constants.
Physical boundary conditions are imposed on the  wave function and  single-valuedness  implies that $M$ can only take integer values $M=0,\pm 1, \pm 2 , \cdots$, but normalisability further restrict $M$ to non-negative values only ($M\ge 0$).
Therefore the normalised  solutions of the equation $a_L^{\phantom{\dag}}\,\psi_L^{(1)}=0 $ are:
\begin{equation}
\psi_{L}^{(1)} \,=\, u_{0,M}\,=\,  C\,  r^M \, e^{-\frac{\nu}{2\lambda \hbar} r^2}  \, e^{iM\varphi}\,, \qquad \qquad M= 0, 1, 2,\, \cdots\, .
\label{zeroLmodeNORM}
\end{equation}
with $C= (\nu/(\lambda\hbar))^\frac{M+1}{2} / (\sqrt{\pi}\, \Gamma(M+1))$.

A spinor solution of the Hamiltonian equation with the upper component $\psi_{L,0}^{(1)}$ given by  Eq.~\eqref{zeroLmodeNORM} can be associated with a null lower component ($\psi_{L,0}^{(2)}=0$) to build a (singlet like) spinor state  solution of the full Hamiltonian eigenvalue equation $H\psi=E\psi$ with energy $E=+mc^2$ for which again \emph{both} intertwining solutions are satisfied:
\beq
(\eps_-=0)\qquad E=+mc^2,\qquad \psi_{L,0}=\left(\ba{c} u_{0,M}\\ 0 \ea\right)\quad  M= 0, 1, 2,\, \cdots\, . \quad (\text{\small{solution of Eqs.~\eqref{eigen}}} )\,.\label{zeroLreg}
\eeq
This is an acceptable  (normalized) solution of our eigenvalue problem. Note that this zero mode has an infinite degeneracy with respect to the positive values of $M$,  -- also note that here, and thereafter, by zero modes we mean states that are annihilated by the corresponding annihilation operators entering the problem, and singlet (doublet) are referred to spinors with only one (both)  component(s) non-vanishing--.

The second of Eqs.~\eqref{zeroLmodes} can be solved in a similar way and we do not give the details here. It is easily found that the  solution  $\psi_L^{(2)}= C\, r^{-M}\, e^{+\frac{\nu}{2\lambda \hbar} r^2}\, e^{iM\varphi} $   can be made regular in the origin by choosing negative values of $M$ but it is always diverging at large radial distances ($r\to \infty$). While one would still  be able to construct a spinor (singlet) solution of the hamiltonian eigenvalue equations, c.f. Eqs.~\eqref{zeroLmodes},  corresponding to $E=-mc^2$ or ($\eps_+ =0$) this would be un-normalizable for any value of $M$ and hence unphysical and  is therefore discarded.



We now turn to a detailed discussion of the  excited states and their energy eigen-values which  are best discussed by considering the second order eigenvalue equation $H^2 \psi_L = E^2 \psi_L $ which allows to disentangle the two components $\psi_L^{(1)}$ and $\psi_L^{(2)}$. Then form Eq.~\eqref{h3} we obtain the decoupled equations: 
\begin{subequations}\label{Pi2}
\begin{align}
{\hat\Pi}_-\, {\hat\Pi}_+\, \psi_{L}^{(1)} &= \epsilon_+\epsilon_- \, \psi_{L}^{(1)}\,,\label{Pi21}\\
{\hat\Pi}_+\, {\hat\Pi}_-\, \psi_{L}^{(2)} &= \epsilon_+\epsilon_- \, \psi_{L}^{(2)}\,,
\label{Pi22}\end{align}
\end{subequations}
where $\eps_+\eps_- = (E^2 -m^2 c^4)/c^2$. By using the explicit form of the creation and annihilation operators given in Eqs.~\eqref{aadag1},\eqref{alr} we find:
\begin{subequations}\label{H2L}
\begin{align}
   \, a^\dag_L \,a^{\phantom{\dag}}_L \,&=\, \frac{ {\hat\Pi}_-\, {\hat\Pi}_+}{4\nu\lambda\hbar}\, =\,  \frac{1}{2\hbar}  
\left[  {H}^\text{2D}_\circ-L_z -\hbar  \right]\,,
\label{H2L1}\\
a^{\phantom{\dag}}_L\, a^\dag_L\,&=\, \frac{{\hat\Pi}_+\, {\hat\Pi}_-}{4\nu\lambda\hbar}\, = \,    \frac{1}{2\hbar} 
\left[ {H}^\text{2D}_\circ -L_z +\hbar \right]\,,\label{H2L2}
\end{align}
\end{subequations}
where we have indicated  the angular momentum in the two dimensional commuting plane $(x,y)$ by the operator $L_z= x\,p_y -y\,p_x$ and:
\begin{equation}
\label{2Dcirc}
{H}^\text{2D}_\circ \,=\,\frac{\lambda}{2\nu} \left( p_x^2+p_y^2\right) + \frac{\nu}{2\lambda} (x^2+y^2)\,.
\end{equation}
 We see therefore that the second order Hamiltonians in Eqs.~\eqref{Pi2}\footnote{Note also that Eqs.~\eqref{H2L} prove indirectly the relation $[a^{\phantom{\dag}}_L\, ,a^\dag_L ] =1$.} can be related to ${H}^\text{2D}_\circ$, the Hamiltonian of a well known \emph{and exactly solvable} non-relativistic system --that of a two dimensional isotropic (or circular) harmonic  oscillator (of unit frequency and mass $\nu/\lambda$)--. The eigen-functions and eigenvalues of this system are well known~\cite{Flugge:1974aa} and can be readily used to solve the second order Hamiltonians of Eqs.~\eqref{Pi2} since the angular momentum operator $L_z$ commutes  with ${H}^{\text{2D}}_\circ$.    
The complete set of eigenfunctions of ${H}^{\text{2D}}_\circ$ and the corresponding eigenvalues  are conveniently derived using the polar  coordinates of the $(x,y)$  plane and they are identified  by a radial quantum number $n_r =0, 1, 2, \cdots$ and the angular momentum quantum number $M=0,\pm1, \pm2, \cdots$~\cite{Flugge:1974aa}:
\begin{equation}
{H}^{\text{2D}}_\circ\,  u_{n_r,M}(r,\varphi)\, = \, {\varepsilon}\,   u_{n_r,M}(r,\varphi) \,, 
\end{equation}
whose eigenstates and eigenvalues are given by:
\begin{subequations}\label{unrM}
\begin{align}
{\varepsilon} \,&=\,  \hbar \left( |M| + 1 + 2n_r \right)\,,\\
u_{n_r,M}(r,\varphi)\, &= \, C_{n_r,M} \, r^{|M|} \, e^{-\frac{\nu}{2\lambda \hbar} r^2}\! \!\!\phantom{F}_1F_1 (-n_r, |M|+1; \frac{\nu}{\lambda\hbar}r^2) \, e^{iM\varphi}\,,
\end{align}
\end{subequations}
where $C_{n_r,M}$ are normalisation constants that can be easily computed as:
\begin{equation}
\label{Cnorm}
C_{n_r,M}\, = \,
\frac{1}{\sqrt{\pi}} \, \left( \frac{\nu}{\lambda \hbar}\right)^{\frac{|M|+1}{2}}\, \frac{\sqrt{\Gamma(|M| +1 +n_r)}}{\Gamma(|M|+1)\sqrt{\Gamma(n_r+1)}}\,.
\end{equation}
The relativistic eigenvalues of the second order equations  are then easily found. The solution of  Eq.~\eqref{Pi21} is obtained with the help of  Eq.~\eqref{H2L1} and is found to be:
\begin{eqnarray}
\psi_L^{(1)}(r,\varphi) &=& C_1 \, u_{n_r,M}(r,\varphi)\,,\\
\eps_+\eps_- \, &=& \, 4 \nu \lambda \hbar\,  \frac{1}{2\hbar} \, \left[ \hbar \left(  |M| + 1 + 2n_r \right)-\hbar M -\hbar\right]\,, 
\end{eqnarray}
where $C_1$ is a spinor normalisation constant. From the above equation the excited states eigenvalues ($n_r\ge 1$) of the original Dirac equation can be extracted:
\begin{equation}
\label{spectrumL1}
E_N^{\pm}=\pm mc^2\sqrt{1+4\, \frac{\lambda\nu\hbar}{m^2c^2}\, N }\, =\, 
\pm mc^2\sqrt{1+\,\zeta_L\, N }\,,\quad   N=n_r+\frac{|M|-M}{2}\quad N=1,2,\,,\cdots\,,
\end{equation}
and the quantity $\zeta_L$ is defined as:
\begin{equation}
\label{spectrumL2}
\zeta_L=\zeta_L(\tilde{\omega}-\omega;\omega_\eta,\omega_\theta)\,=\,\frac{4\lambda \nu \hbar}{m^2c^2}\, =\, -4\,\frac{\hbar}{mc^2}\, \left(1-\frac{\tilde{\omega}-\omega}{\omega_\theta}\right)(\tilde{\omega}-\omega-\omega_\eta)\,.
\end{equation}

We see that every energy level $E_N$ is highly degenerate.  In particular every level has an infinite degeneracy with respect to the non-negative values of $M$, ($M\ge 0$), while there is a finite degeneracy $D=N+1$ with respect to the negative values of $M$. 

The lower component $\psi^{(2)}_L$ of the eigen-solution of the Dirac Hamiltonian is found by using the intertwining relation in Eq.~\eqref{eigen}:
\begin{eqnarray}
\label{second.comp}
\psi^{(2)}_L (r,\varphi)\,&=&\, - 2i \frac{\sqrt{\lambda \nu \hbar}}{\eps_+} \, a_L^{\phantom{\dag}}\, C_1\,  u_{n_r,M}(r,\varphi)\,,\nonumber\\
&=& -i \frac{C_1}{\eps_+} \, e^{i\varphi}\, \left\{\lambda \hbar \left(\partial_r +\frac{i}{r}\partial_\varphi\right)+\nu r\right\}\,u_{n_r,M}(r,\varphi) \,.
\end{eqnarray}
We find that the sign of the angular quantum number ($M$) identifies two classes of eigen-states which we discuss separately.

\vspace{0.1cm}\noindent \underline{\bm{$M \ge 0$} (inifinite degeneracy)}:\\
In this case using the explicit expression of the wave functions $u_{n_r,M}$ and with the aid of the recurrence relation~\cite{Abramowitz:1964aa}:
\begin{equation}
\frac{d}{dz} \! \!\!\phantom{F}_1F_1  (a,b;z) = \frac{a}{b} \! \!\!\phantom{F}_1F_1  (a+1,b+1;z)\,,
\end{equation}
and after taking into account that for $M\ge 0$ the normalisation constants in Eq.~\eqref{Cnorm} satisfy $C_{n_r,M} = C_{n_r-1,M+1}\,\sqrt{\lambda \hbar/\nu}\,(M+1)/\sqrt{n_r}$ we find explicitly:
\begin{equation}
\psi^{(2)}_L \,=+ i \, C_1 \frac{\sqrt{4\nu\lambda\hbar n_r}}{\eps_+}\, u_{n_r-1,M+1} =  \, + i \, C_1 \,\frac{\sqrt{\eps_- \eps_+}}{\eps_+}\, \, u_{n_r-1,M+1}\,.
\end{equation}

Finally the normalised spinor solution for positive values of $M$ can be put in the form:
\begin{equation}
D=\infty,~~~~\psi_{L}^{(\pm,n_r,M)}=\, \frac{1}{\sqrt{2}}\, \left(\ba{l} \phantom{\pm \, i\,}\sqrt{1+\frac{mc^2}{E^{\pm}_{N}}} \, u_{n_r,M} \\  \pm \, i\, \sqrt{1-\frac{mc^2}{E^{\pm}_{N}}}\, u_{n_r-1,M+1}\ea\right),~~~~N=n_r=1,2,\, \cdots \, ,
\eeq
where the upper (lower) sign corresponds respectively to the positive  (negative) branch of the spectrum, {and the normalisation condition is, here and thereafter, the usual one for two component spinors, i.e. $\langle\psi|\psi\rangle= \langle\psi_1|\psi_1\rangle+\langle\psi_2|\psi_2\rangle = 1$}.

\vspace{0.2cm}\noindent \underline{\bm{$M < 0$} (finite degeneracy)}:

In this  case when solving for the lower component of the spinor solution in Eq.~\eqref{second.comp} one has to use a different recurrence relation satisfied by the confluent hypergeometric function~\cite{Abramowitz:1964aa}:
\begin{equation}
z \frac{d}{dz} \! \!\!\phantom{F}_1F_1  (a,b;z) = (b-1)\left[ \! \!\!\phantom{F}_1F_1  (a,b-1;z)- \! \!\!\phantom{F}_1F_1 (a,b;z)\right]\,,
\end{equation}
and after taking into account that for $M <0 $ the normalisation constants in Eq.~\eqref{Cnorm} satisfy $C_{n_r,M} = -C_{n_r,M+1}\,\sqrt{\nu\,(n_r-M)/(\lambda \hbar)}\,/M $ we find explicitly:
\begin{equation}
\psi^{(2)}_L \,=- i \, C_1 \frac{\sqrt{4\nu\lambda\hbar (n_r-M)}}{\eps_+}\, u_{n_r,M+1} =  \, - i \, C_1 \,\frac{\sqrt{\eps_- \eps_+}}{\eps_+}\, u_{n_r,M+1}\,,
\end{equation}
and the final expression of the spinor solution for the $N$-th energy level in the case of negative values of $M$ is:
\begin{equation}
 D=N+1,~~~~\psi_{L}^{(\pm,n_r,M)}=\, \frac{1}{\sqrt{2}}\, \left(\ba{l} \phantom{\pm \, i\,}\sqrt{1+\frac{mc^2}{E^{\pm}_{N}}} \, u_{n_r,M} \\  \mp \, i\, \sqrt{1-\frac{mc^2}{E^{\pm}_{N}}}\, u_{n_r,M+1}\ea\right),~~~~N=n_r - M=1,2,\, \cdots \, ,
\end{equation}
where again the upper (lower) sign correspond to the positive (negative) branch  of the spectrum.


Also we note the energy gap is dependent on the commutative parameters  $\theta, \eta$. 
\section{Intermediate Magnetic Field: ${B_{\text{cr}} < B < B_{\text{cr}}^*}$  (${\omega_\eta <\tilde{\omega} -\omega  < \omega_\theta}$)} \label{inter}

\noindent In this case we find from Eq.~\eqref{pi}: 
\beq
\label{pi2}
{\hat\Pi}_\pm=\lambda(p_x\pm i\,p_y)\, \pm \, i\mu (x \pm i\,y)\,,
\end{equation}
\begin{equation}
\label{paramR}
\lambda= 1- \displaystyle\frac{\tilde{\omega}-\omega}{\omega_\theta} > 0 \,,\qquad \qquad \mu= m (\tilde{\omega} -\omega -\omega_\eta ) > 0\,.
\eeq
We define the following set of \emph{new} creation and annihilation operators: 
\beq\label{aadag}
\ba{lcl}
\displaystyle{\tilde{a}^{\phantom{\dag}}_x=\f{i}{\sqrt{2\mu\lambda\hbar}}(\lambda p_x-i\mu x),~~~~\tilde{a}_x^\dag=\f{-i}{\sqrt{2\lambda\mu\hbar}}(\lambda p_x+i\mu x})\,,\\
\displaystyle{\tilde{a}^{\phantom{\dag}}_y=\f{i}{\sqrt{2\mu \lambda \hbar}}(\lambda p_y-i\mu y),~~~~\tilde{a}_y^\dag=\f{-i}{\sqrt{2\lambda\mu\hbar}}(\lambda p_y+i\mu y})\,,
\ea
\eeq
Left and right creation and annihilation operators, $\tilde{a}^{\phantom{\dag}}_L,\tilde{a}^{\phantom{\dag}}_R, \tilde{a}^{\dag}_L,\tilde{a}^{\dag}_R$ are defined, similarly to Eq.~\eqref{alr}.
In terms of the new creation and annihilation operators  it follows that $\Pi_{\pm}$
are given by:
\begin{equation}
{\hat\Pi}_+=+2\, i\, \sqrt{\mu \lambda\hbar}\, \tilde{a}_R^\dag,~~~~{\hat\Pi}_-=-2\, i\, \sqrt{\mu \lambda\hbar}\, \tilde{a}^{\phantom{\dag}}_R,~~~~\mu>0\,. 
\end{equation}
We see that the Dirac Hamiltonian is in this case expressed solely in terms of the right handed creation and annihilation operators so in this region of parameters the system is said to be in the right chiral phase. 
The component eigenvalue equations corresponding to the Dirac Hamiltonian, c.f. in Eq~\eqref{h3} can be written as:
\beq\label{eigen1}
-2\, i\, \sqrt{\mu \lambda \hbar}\,\,\tilde{a}^{\phantom{\dag}}_R\, \psi^{(2)}_{R}=\eps_-\, \psi^{(1)}_{R},~~~~+2\,i\,\sqrt{\mu \lambda\hbar}\,\,\tilde{a}_R^\dag\, \psi^{(1)}_{R}=\eps_+\, \psi^{(2)}_{R},~~~~\eps_\pm=\f{E\pm mc^2}{c}\, .
\eeq
We again make use of the polar coordinates $r,\phi$ as defined previously and find the explicit expression for the right chiral annihilation and creation operators:
\begin{eqnarray}
\tilde{a}_R^{\phantom{\dag}} \,& =&\, \frac{e^{-i\varphi}}{2\sqrt{\mu\lambda \hbar}} \,\left\{ \phantom{+} \lambda \hbar \, \left( \partial_r -\frac{i}{r}\, \partial_{\varphi} \right) +\mu r \right\}\,,\nonumber \\
\tilde{a}_R^{{\dag}} \, &=&\, \frac{e^{+i\varphi}}{2\sqrt{\mu\lambda \hbar}} \,\left\{ - \lambda \hbar\,\left( \partial_r +\frac{i}{r} \, \partial_{\varphi} \right) +\mu r \right\}\,.\nonumber
\end{eqnarray}
The analysis of the spectrum proceeds in a similar fashion to what has been discussed in detail for the left phase. We skip therefore  the details and give only the final results.

As regards the zero modes in the right phase we find that the normalisable solution is given in terms of the eigenfunction $\tilde{u}_{0,M} = C\, r^{-M}\, e^{-\frac{\mu}{2\lambda\hbar} r^2}$  and now it is infinitely degenerate with respect to the non-positive values of $M$, ( $M\le0$), and is a spin down singlet with $E=-mc^2$ (negative branch of the spectrum): 
\beq
(\eps_+=0)\quad E=-mc^2,\quad \psi_{R,0}=\left(\ba{c} 0 \\ \tilde{u}_{0,M}\ea\right),\quad
 M=0,-1,-2\, \cdots\,, \quad (\text{\small solution of Eqs.~\eqref{eigen1}})\,,\label{zeroRreg}
\eeq 
which is normalized to unity if the constant $C$ is chosen as:\\ $C= (\mu/(\lambda\hbar))^\frac{|M|+1}{2} / (\sqrt{\pi}\, \Gamma(|M|+1))$.

We now describe the excited states. The decoupled equations in the right phase are  obtained from Eq.~\eqref{eigen1}:
\beq
4\mu \lambda\hbar\,  \tilde{a}_R^\dag\, \tilde{a}^{\phantom{\dag}}_R\, \psi_R^{(2)}=\eps_+\eps_-\,\psi_R^{(2)},\qquad \qquad 4\mu\lambda \hbar\, \tilde{a}^{\phantom{\dag}}_R\, \tilde{a}_R^\dag\, \psi_R^{(1)}=\eps_+\eps_-\,\psi_R^{(1)}\,,\label{osci}
\eeq
and their solution is discussed again by computing the second order operators as:
\begin{subequations}\label{H2R}
\begin{align}
   \, \tilde{a}^\dag_R \,\tilde{a}^{\phantom{\dag}}_R \,&=\, \frac{ {\hat\Pi}_+\, {\hat\Pi}_-}{4\mu\lambda\hbar}\, =\,  \frac{1}{2\hbar}  
\left[  \widetilde{H}^\text{2D}_\circ+L_z -\hbar  \right]\,,
\label{H2R1}\\
\tilde{a}^{\phantom{\dag}}_R\, \tilde{a}^\dag_R\,&=\, \frac{{\hat\Pi}_-\, {\hat\Pi}_+}{4\mu\lambda\hbar}\, = \,    \frac{1}{2\hbar} 
\left[ \widetilde{H}^\text{2D}_\circ +L_z +\hbar \right]\,,\label{H2R2}
\end{align}
\end{subequations}
where $\widetilde{H}^\text{2D}_\circ$ is a non relativistic Hamiltonian of a circular oscillator of unit frequency and mass $\mu/\lambda$:
\begin{equation}
\label{2DcircR}
\widetilde{H}^\text{2D}_\circ \,=\,\frac{\lambda}{2\mu} \left( p_x^2+p_y^2\right) + \frac{\mu}{2\lambda} (x^2+y^2)\,,
\end{equation}
which differs form ${H}^\text{2D}_\circ$ in Eq.~\eqref{2Dcirc} only by the change 
$\nu \to \mu$. Note that this is the only change that will affect the eigenfunctions of $\widetilde{H}^\text{2D}_\circ$ ($\tilde{u}_{n_r,M}$) and their normalisation constants $\widetilde{C}_{n_r,M}$ with respect to those of ${H}^\text{2D}_\circ$ (${u_{n_r,M}}$ and ${C}_{n_r,M}$), c.f. Eqs.~(\ref{unrM},\ref{Cnorm}).   
We only mention here that  in Eqs.~\eqref{H2R} the sign of the angular momentum operator is changed relative to Eqs.\eqref{H2L} which anticipates that in the right phase the role of the angular momentum quantum number is reversed. Therefore in the right phase we anticipate the following behaviour: infinite degeneracy for $M\le 0$ and finite degeneracy for positive values of $M$ ($M>0$) i.e. opposite to the what happens in the left phase. 

The excited levels ($n_r\ge 1$) are then derived from the second order equations c.f. Eq~\eqref{Pi2} similarly to what has been done in the left chiral phase. We find it convenient to solve the lower component $\psi_R^{(2)}$ from Eq.~\eqref{Pi22} and then compute the upper component $\psi_R^{(1)}$ with the first of the intertwining relations in Eq.~\eqref{eigen1}. We find:
\begin{equation}
E_N^{\pm}=\pm mc^2\sqrt{1+4\, \frac{\lambda\mu\hbar}{m^2c^2}\, N }\, =\, 
\pm mc^2\sqrt{1+\,\zeta_R\, N }\qquad   N=n_r+\frac{M+|M|}{2}\qquad N=1,2 \cdots
\end{equation}
and the quantity $\zeta_R$ is defined as:
\begin{equation}
\label{spectrumR}
\zeta_R=\,\zeta_R(\tilde{\omega}-\omega;\omega_\eta,\omega_\theta)\,=\,4\,\frac{ \mu \lambda \hbar}{m^2c^2}\, =\, 4\,\frac{\hbar}{mc^2}\, \left(1-\frac{\tilde{\omega}-{\omega}}{\omega_\theta}\right)(\tilde{\omega}-\omega-\omega_\eta)\,.
\end{equation}
We see again that every energy level $E_N$ is highly degenerate and, as expected, there is an infinite degeneracy with respect to the non-positive values of $M$, ($M \le 0$), while the degeneracy is finite, $D=N+1$, with respect to the positive values of $M$. The corresponding eigen-solutions are:
\\ 

\noindent \underline{\bm{$M > 0$} (finite degeneracy)}:\\
In this case using the explicit expression of the wave functions $\tilde{u}_{n_r,M}$ 
we find the normalised spinor solution for positive values of $M$ can be put in the form:
\begin{equation}
D=N+1,~~~~\psi_{R}^{(\pm,n_r,M)}=\, \frac{1}{\sqrt{2}}\, \left(\ba{l} {\mp \, i\,}\sqrt{1+\frac{mc^2}{E^{\pm}_{N}}} \, \tilde{u}_{n_r,M-1} \\  \phantom{\pm \, i\,} \sqrt{1-\frac{mc^2}{E^{\pm}_{N}}}\, \tilde{u}_{n_r,M}\ea\right),~~~~N=n_r+M=1,2,\,\cdots\, ,
\eeq
where the upper (lower) sign corresponds respectively to the positive (negative) branch of the spectrum.\\ 

\noindent \underline{\bm{$M \le 0$} (infinite degeneracy)}:
\begin{equation}
D=\infty,~~~~\psi_{R}^{(\pm,n_r,M)}=\, \frac{1}{\sqrt{2}}\, \left(\ba{l} {\pm \, i\,}\sqrt{1+\frac{mc^2}{E^{\pm}_{N}}} \, \tilde{u}_{n_r-1,M-1} \\  \phantom{\pm \, i\,} \sqrt{1-\frac{mc^2}{E^{\pm}_{N}}}\, \tilde{u}_{n_r,M}\ea\right),~~~~N=n_r=1,2,\, \cdots\,.
\eeq

Thus for $B_{\text{cr}} < B < B_{\text{cr}}^*$, the $E={-}mc^2$ state is a singlet while for the other energy values the positive and negative energy states are paired in doublets. It may also be noticed that the energy gap is not $2mc^2$ but depends  on the magnetic field intensity $B$ as well as on the  non-commutativity parameters $\eta$ and $\theta$.\\

\section{\label{strong} Strong Magnetic Field: ${B > B^*_\text{cr}}$ (${\tilde{\omega} -\omega >  \omega_\theta}$).} 

\noindent 
We see that relative to case 2 (see Eq.~\eqref{pi2}) now the quantity  $\lambda =1-( \tilde{\omega}-\omega )/\omega_\theta$ becomes negative and therefore the operators ${\hat\Pi}_\pm$ are given by
\begin{equation}
\label{PiLp}
{\hat\Pi}_\pm =- \left[ \delta(p_x\pm i\,p_y)\mp i\mu (x \pm i\,y) \right]\,,
\end{equation}
where:
\begin{equation}
\delta= \displaystyle\frac{\tilde{\omega}-\omega}{\omega_\theta} -1> 0 \,,
\end{equation}
and $\mu$ is given as in Eq.~\ref{paramR}
and we see that $\Pi_\pm$ in Eq.~\eqref{PiLp} differ by an overall negative sign with the operators in Eq.~\eqref{pi1}, of case 1 (the region with a weak magnetic field), --the so called left chiral phase--  with the substitutions: $\lambda \to \delta, \nu \to \mu$. We expect therefore this region of very strong magnetic field ($\tilde{\omega} -\omega >  \omega_\theta$) to be  described as left chiral phase.

This is easily shown by defining still another set of  creation and annihilation operators: 
\beq\label{aadag}
\ba{lcl}
\displaystyle{b^{\phantom{\dag}}_x=\f{i}{\sqrt{2\mu\delta\hbar}}(\delta p_x-i\mu x),~~~~b_x^\dag=\f{-i}{\sqrt{2\mu\delta\hbar}}(\delta p_x+i\mu x})\,,\\
\displaystyle{b^{\phantom{\dag}}_y=\f{i}{\sqrt{2\mu \delta \hbar}}(\delta p_y-i\mu y),~~~~b_y^\dag=\f{-i}{\sqrt{2\mu\delta\hbar}}(\delta p_y+i\mu y)}\,.
\ea
\eeq
Left and right creation and annihilation operators, $b^{\phantom{\dag}}_L,b^{\phantom{\dag}}_R, b^{\dag}_L,b^{\dag}_R$ are defined, similarly to Eq.~\eqref{alr}.
In terms of the new creation and annihilation operators  it follows that $\Pi_{\pm}$
are given by:
\begin{equation}
\label{LII}
{\hat\Pi}_+=+\,2\, i \, \sqrt{\delta \mu \hbar}\, b^{\phantom{\dag}}_L,\qquad \qquad{\hat\Pi}_-= -\,2\,i\,\sqrt{\delta\mu \hbar}\,b_L^\dag ,\qquad\qquad\delta,\mu>0 \,.
\end{equation}
We see therefore that the system, apart from an overall sign in the definition of the operators $\hat{\Pi}_\pm$, is equivalent to that described in Eq.~\eqref{LI} characterised by left chiral excitations. Evidently the second order equations are going to be identical except for the replacements of the relevant parameters $\lambda \to \delta, \nu \to \mu$.  

The intertwining relations, corresponding to the (first-order) Dirac Hamiltonian eigenvalue equation, are therefore: 
\beq
-2\, i\, \sqrt{\mu\delta\hbar}\, b_L^\dag \, \psi_L^{(2)}=\eps_-\psi_L^{(1)},~~~~ +2\, i\, \sqrt{\mu\delta\hbar}\, b_L^{\phantom{\dag}}\, \psi_L^{(1)}=\eps_+\psi_L^{(2)},~~~~\eps_\pm=\f{E\pm mc^2}{c}\,,\label{eigenLp}
\eeq
which apart from the replacement of the parameters differ only by a sign with those of case 1.

The  annihilation and creation operators in standard polar coordinates are now given by:
\begin{eqnarray}
b_L^{\phantom{\dag}} \,& =&\, \frac{e^{i\varphi}}{2\sqrt{\mu\delta \hbar}} \,\left\{ \phantom{+} \delta \hbar \, \left( \partial_r +\frac{i}{r}\, \partial_{\varphi} \right) +\mu r \right\}\,,\nonumber \\
b_L^{{\dag}} \, &=&\, \frac{e^{-i\varphi}}{2\sqrt{\mu\delta \hbar}} \,\left\{ - \delta \hbar\,\left( \partial_r -\frac{i}{r} \, \partial_{\varphi} \right) +\mu r \right\}\,.\nonumber
\end{eqnarray}
We examine first whether there are solutions  with energies $E=\pm mc^2$ i.e, $\eps_{\pm}=0$,
and find now that the normalisable solution is given in terms of the eigenfunction ${u'}_{0,M} = C'\, r^{M}\, e^{-\frac{\mu}{2\delta\hbar} r^2}$,  infinitely degenerate with respect to the non-negative values of $M$, ($M\ge0$), and is a spin-up singlet with $E=+mc^2$ (positive branch of the spectrum): 
\beq
(\eps_-=0)\qquad E=+mc^2,\qquad \psi_{R,0}=\left(\ba{c} {u'}_{0,M} \\ 0\ea\right),\quad
M=0,1,2\, \cdots\,, \quad (\text{\small{solution of Eqs.~\eqref{eigenLp}}})\label{zeroLpreg}
\eeq 
which is normalized to unity if the constant $C'$ is chosen as: \\$C'= (\mu/(\delta\hbar))^\frac{M+1}{2} / (\sqrt{\pi}\, \Gamma(M+1))$.

We now describe the excited states, and their solution is discussed again by computing the second order operators as:
\begin{subequations}\label{H2Lp}
\begin{align}
   \, \tilde{b}^\dag_L \,\tilde{b}^{\phantom{\dag}}_L \,&=\, \frac{ {\hat\Pi}_-\, {\hat\Pi}_+}{4\mu\delta\hbar}\, =\,  \frac{1}{2\hbar}  
\left[  {H'}^\text{2D}_\circ-L_z -\hbar  \right]\,,
\label{H2Lp1}\\
\tilde{b}^{\phantom{\dag}}_L\, \tilde{b}^\dag_L\,&=\, \frac{{\hat\Pi}_+\, {\hat\Pi}_-}{4\mu\delta\hbar}\, = \,    \frac{1}{2\hbar} 
\left[ {H'}^\text{2D}_\circ -L_z +\hbar \right]\,,\label{H2Lp2}
\end{align}
\end{subequations}
where ${H'}^\text{2D}_\circ$ is a non relativistic Hamiltonian of a circular oscillator of unit frequency and mass $\mu/\delta$:
\begin{equation}
\label{2DcircR}
\widetilde{H}^\text{2D}_\circ \,=\,\frac{\delta}{2\mu} \left( p_x^2+p_y^2\right) + \frac{\mu}{2\delta} (x^2+y^2)\,,
\end{equation}
which differs form ${H}^\text{2D}_\circ$ in Eq.~\eqref{2Dcirc} only by the exchanges 
$(\nu \to \mu, \lambda \to \delta)$. Note that this is the only change that will affect the eigenfunctions of ${H'}^\text{2D}_\circ$ (${u'}_{n_r,M}$) and their normalisation constants ${C'}_{n_r,M}$ with respect to those of ${H}^\text{2D}_\circ$ (${u_{n_r,M}}$ and ${C}_{n_r,M}$), c.f. Eqs.~(\ref{unrM},\ref{Cnorm}).   
We only mention here that  in Eqs.~\eqref{H2Lp} the sign of the angular momentum operator is changed relative to Eqs.~\eqref{H2R} and is again the same as that of Eqs.~\eqref{H2L} which shows that the role of the angular momentum quantum number is once again reversed and now it is identical to that of case 1. 
The excited levels ($n_r\ge 1$) are then derived from the second order equations c.f. Eq~\eqref{Pi2} in exactly the same way as done in the left chiral phase (case 1).

The spectrum and the eigenfunctions are then:
\begin{equation}
\label{spectrumLp1}
E_N^{\pm}=\pm mc^2\sqrt{1+4\, \frac{\delta\mu\hbar}{m^2c^2}\, N }\, =\, 
\pm mc^2\sqrt{1+\,\zeta_L'\, N }\qquad   N=n_r+\frac{|M|-M}{2}\qquad N=1,2,\cdots
\end{equation}
and the quantity $\zeta_L'$ is defined by:
\begin{equation}
\label{spectrumLp2}
\zeta_L'=\zeta_L'(\tilde{\omega}-\omega;\omega_\eta,\omega_\theta)\,=\,\frac{4\,\delta \mu \,\hbar}{m^2c^2}\, =\, 4\,\frac{\hbar}{mc^2}\, \left(\frac{\tilde{\omega}-\omega}{\omega_\theta}-1\right)(\tilde{\omega}-\omega-\omega_\eta)\,.
\end{equation}
\noindent \underline{\bm{$M \ge 0$} (infinite degeneracy)}:\\
The normalised spinor solution for positive values of $M$ can be put in the form:
\begin{equation}
D=\infty,~~~~\psi_{L}^{(\pm,n_r,M)}=\, \frac{1}{\sqrt{2}}\, \left(\ba{l} \phantom{\pm \, i\,}\sqrt{1+\frac{mc^2}{E^{\pm}_{N}}} \, u_{n_r,M} \\  \mp \, i\, \sqrt{1-\frac{mc^2}{E^{\pm}_{N}}}\, u_{n_r-1,M+1}\ea\right),~~~~N=n_r=1,2,\, \cdots\,,
\eeq
where the upper (lower) sign corresponds respectively to the positive  (negative) branch of the spectrum.\\

\noindent \underline{\bm{$M < 0$} (finite degeneracy)}:\\
The final expression of the spinor solution for the $N$-th energy level in the case of negative values of $M$ is:
\begin{equation}
 D=N+1,~~~~\psi_{L}^{(\pm,n_r,M)}=\, \frac{1}{\sqrt{2}}\, \left(\ba{l} \phantom{\pm \, i\,}\sqrt{1+\frac{mc^2}{E^{\pm}_{N}}} \, u_{n_r,M} \\  \pm \, i\, \sqrt{1-\frac{mc^2}{E^{\pm}_{N}}}\, u_{n_r,M+1}\ea\right),~~~~N=n_r - M=1,2,\cdots\,,
\end{equation}
where again the upper (lower) sign correspond to the positive (negative) branch  of the spectrum.

Note that the spinor structure of the eigenstates of the excited levels in this new left chiral phase is identical to that of the initial left phase except for the relative sign between the lower and upper components (in addition of course to the replacement of the parameters).\\  

\begin{figure*}[t!]
\includegraphics[width=16.0cm]{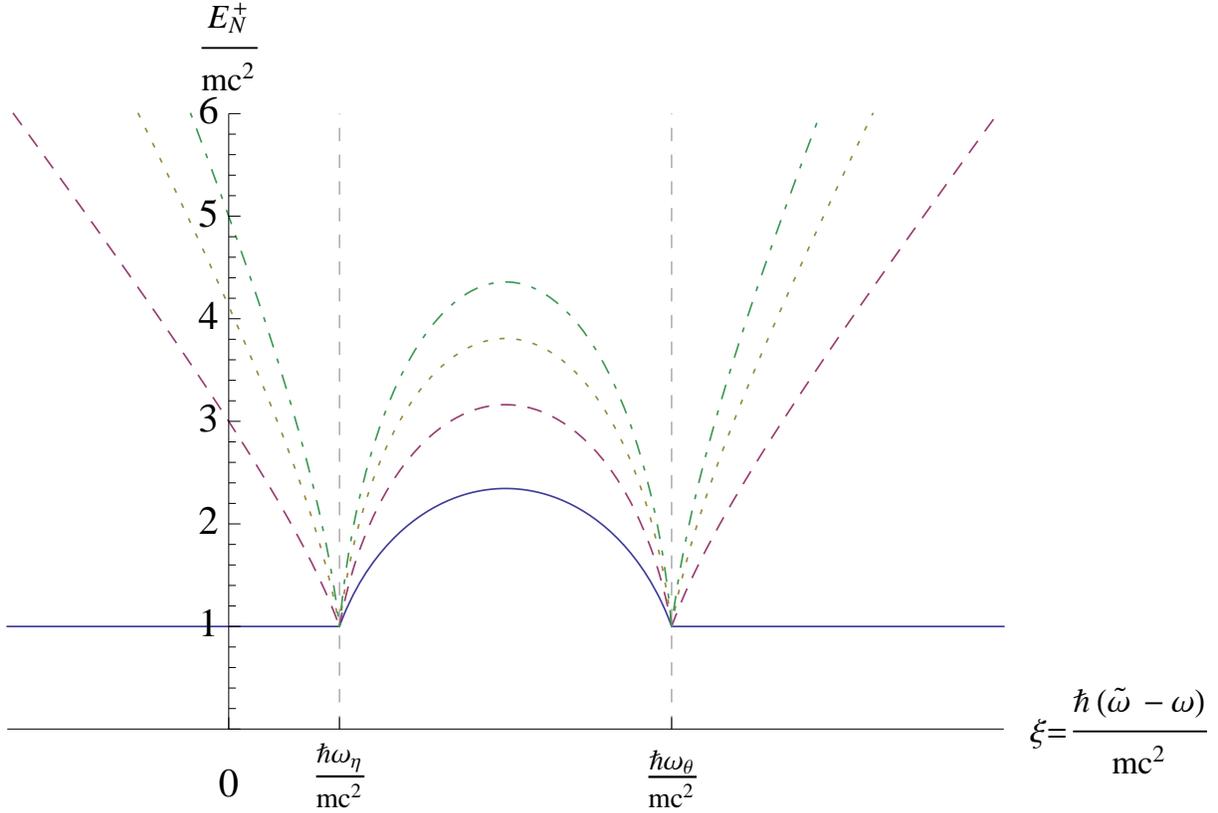}
\caption{\label{fig:fig1} (Color online) Energy eigenvalues of the positive branch $E_N^+$, in units of $mc^2$, of the ground state, solid line (blue)  and of the first few excited states, dashed (red), dotted (light green) and dot-dashed (green), as a function of the dimensionless ratio $\xi=\hbar (\tilde{\omega} -\omega)/(mc^2)$. This is achieved combining the expressions derived in the text in Eqs.~(\ref{spectrumL1},\ref{spectrumR},\ref{spectrumLp2}).   There are two values of   the magnetic field (or $\tilde{\omega}$) for which there is a \emph{quantum} phase transition (a point of non analyticity in the energy eigenvalues). In the commutative limit ($\eta,\theta \to 0$) $\omega_\eta \to 0$ and $\omega_\theta \to \infty$ and  the spectrum goes smoothly into the one discussed in \cite{Bermudez:2008ab} with  only one phase transition at $\xi=0$.
 }
\end{figure*}

\section{\bf{Discussion}}
\label{discussion}
In Fig.~\ref{fig:fig1} we show the positive branch of the spectrum for the first few energy levels as function of the dimensionless quantity $\xi=\hbar(\tilde{\omega}-\omega)/(mc^2)$ which is directly related to the intensity of the external magnetic field through $\tilde{\omega}= eB/(2mc)$. Note that while in the two left phases the positive branch of the spectrum starts with the singlet zero mode at $E=+mc^2$ in the right phase the lowest positive energy state (the ground state) is a doublet with $E=+mc^2\sqrt{1+\zeta_R}$. In Fig.~\ref{fig:fig1} one clearly makes out the two points of  non-analyticity, $\xi=\hbar\omega_\eta/(mc^2)$ and $\xi=\hbar\omega_\theta/(mc^2)$, which signal the two quantum phase transitions. The same behaviour is of course shown by all the excited states. In the commutative limit ($\eta,\theta \to 0$) we have that  $\omega_\eta \to 0$ and $\omega_\theta \to \infty$ and therefore we obtain that while the first phase transition (left-right) is shifted towards  the commutative critical value $\xi=0$ ($\tilde{\omega} =\omega$), the second phase transition (right-left) disappears since it is moved to $\xi=\infty$ ($\tilde{\omega}=\infty$).


Another point worth to be emphasised is related to the definiteness of the average value of the angular momentum, $L_z$, in each of the classes of states of finite or infinite degeneracy as described in the previous   paragraphs. In each one of the chiral phases $(i=L,R,L')$ the average angular momentum $\langle L_z \rangle^{(i)}$  is given by the following compact expression:
\begin{equation}
\label{averageL}
\langle L_z \rangle^{(i)} = \hbar \left[ M+\frac{\alpha_i}{2}\left(1  \mp\, \alpha_i\,\frac{1}{ \sqrt{1+ \zeta_{(i)} N}}\right)\right]  \qquad \text{for\ }  \qquad N=1,2,\, \cdots\,, 
\end{equation}
where: the upper sign is for the positive branch of the energy spectrum while the lower sign holds for the negative branch;  $\alpha_i = 1$ for the two left chiral phases $i=L,L'$ and $\alpha_i=-1$ for the right chiral phase ($i=R$) and the functions $\zeta_{(i)}$ are given in  Eqs.~(\ref{spectrumL1},\ref{spectrumR},\ref{spectrumLp2}). Note that in all cases this expression holds for both classes of finite and infinite degeneracy with respect to the angular quantum number $M$. For instance in the two left phases ($L,L'$) for the states with finite degeneracy ($M<0$) we have that the angular momentum is negative definite, $\langle L_z \rangle < 0 $ while in the class with $M\ge0$ (infinite degeneracy) the average angular momentum is semi-positive definite $\langle L_z \rangle \ge0 $.  This can be easily verified from Eq.~\eqref{averageL} given that in the left-chiral phase we are considering now $\zeta_L \ge 0$, ($\zeta_L$ will vanish when $B=B_{\text{cr}}$). Note that there is a single  exception to this rule. For $M=-1$ the average angular momentum vanishes for the negative branch of the spectrum ($\langle L_z \rangle = 0$) at the critical point $B=B_{\text{cr}}$. Viceversa in the right chiral phase ($R$) for the states with finite degeneracy ($M>0$) we have that the angular momentum is positive definite, $\langle L_z \rangle >0 $ while in the class with $M\le0$ (infinite degeneracy) the average angular momentum is semi-negative definite $\langle L_z \rangle \le0 $.  This can be easily verified from Eq.~\eqref{averageL} given that in the right-chiral phase  $\zeta_R \ge 0$, ($\zeta_R$ will vanish only when $B=B_{\text{cr}}$ and $B=B_{\text{cr}}^*$).
Also in the right phase there is an exception to this rule. For $M=+1$ the average angular momentum vanishes ($\langle L_z \rangle = 0$) at the two critical points $B=B_{\text{cr}}$ and $B=B_{\text{cr}}^*$ in the positive branch. 
Note that the above classification of the solutions in classes of finite/infinite degeneracy, $D$,
each one with an angular momentum average of definite (or semi-definite) sign is independent of the presence of non-commutativity. 

We point out that in previous discussions of the left-right chiral phase transition of the Dirac oscillator in a constant homogeneous magnetic field the presence of the classes of solutions with opposite angular momentum average (with finite degeneracy in the left phases and infinite degeneracy in the right phase) was apparently overlooked~\cite{Quimbay:2013aa,Bermudez:2008ab,Bermudez:2007aa,Bermudez:2007ab,Bermudez:2008aa,Mandal:2010aa,Mandal:2012aa}.  

We now would like to discuss a physical observable,  closely related to the spectrum, the magnetisation, which has the potential of characterising the quantum phase transitions of the system. Indeed for every energy level $E_N^{(i)}$, $i=L,R,L'$, we can define the magnetisation $M_N^{(i)}$ by:
\begin{equation}
M_N^{(i)}= -\frac{\partial E_N^{(i)}}{\partial B}\,.
\end{equation}
In each of the quantum phases that we have discussed previously the energy eigenvalues can be written as $E_N^{(i)}= mc^2 \sqrt{1+\zeta_{(i)} N} $  with the functions $\zeta_{(i)}$ as in Eqs.~(\ref{spectrumL1},\ref{spectrumR},\ref{spectrumLp2}) and the magnetisation is then easily found to be given by:
\begin{eqnarray}
M_N^{(i)}&=& - \frac{mc^2}{2\sqrt{1+N\zeta_{(i)} }}\, N \,\left(\frac{\partial \zeta_{(i)}}{\partial B} \right)\nonumber \\ &=& - 2\, \mu_B\, \frac{m_e}{m} \, \frac{N}{\sqrt{1+N\zeta_{(i)} }} \, \left( 1 + \frac{\omega_\eta}{\omega_\theta} -2\,\frac{\tilde{\omega}-\omega}{\omega_\theta}\right)\,. 
\end{eqnarray}
where $m_e$ is the electron mass and $\mu_B = \frac{e\hbar}{2m_ec}$ is the Bohr magneton.

In Fig.~\ref{fig:fig2} we show the magnetisation of our system assuming $m_e=m$. We find that for every level the magnetisation has a  finite discontinuity at the two critical points. We point out a very interesting feature which is closely tied to the presence of non-commutativity. We clearly see that in the right chiral phase there is a fixed point of vanishing magnetisation ($M_N^{(R)}=0$) at $\tilde{\omega}-\omega = {(\omega_\eta +\omega_\theta)}/{2}$, exactly the midpoint of the two critical points. Clearly  this will remain so even when computing a thermal average of the magnetisation at a given temperature. Since $M_N^{(R)}=0$ for every level its average value, at any given temperature, will also vanish. So this can be taken as a prediction of our model for the system under consideration: \emph{in the presence of non-commutativity there is a fixed point with vanishing magnetisation in the right chiral phase}. 
\begin{figure*}[t!]
\includegraphics[width=16.0cm]{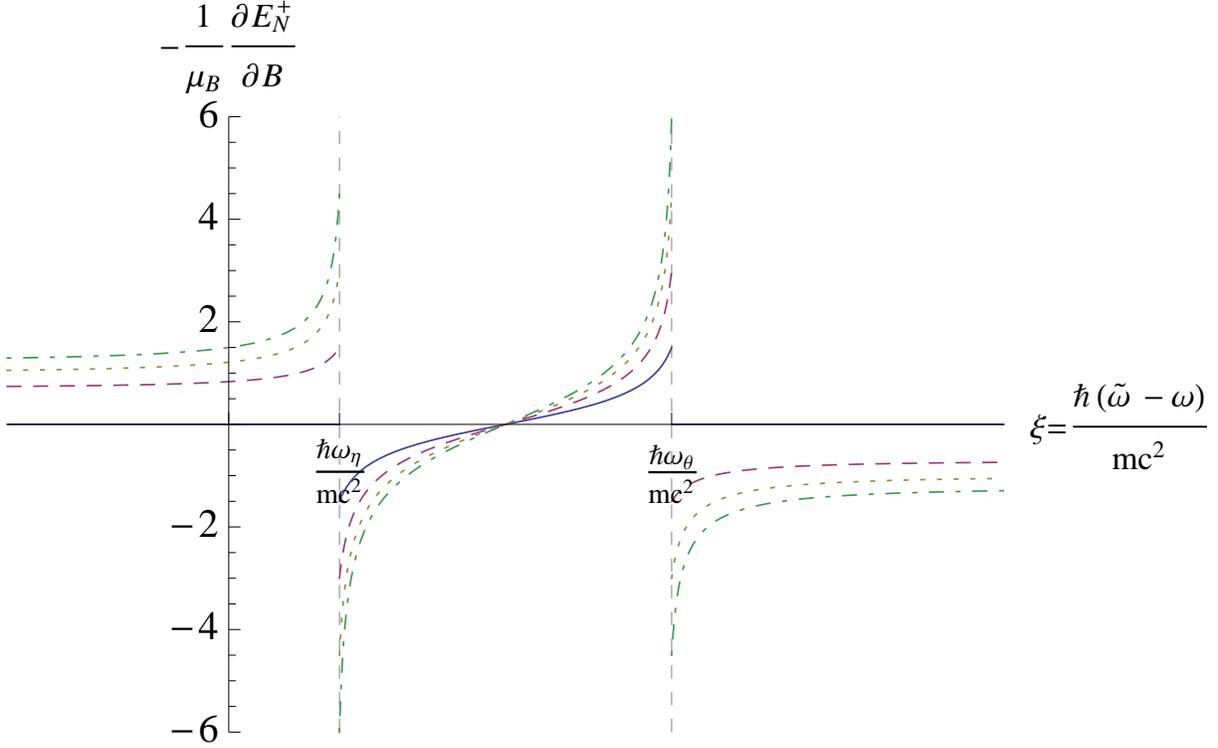}
\caption{\label{fig:fig2} (Color Online) The magnetisation of the energy levels of the positive branch of the spectrum $E_N^+$ is shown for the ground state, solid line (blue) and first few excited energy levels, dashed (red), dotted (light green) and dot-dashed (green), in units of the Bohr magneton $\mu_B=\frac{e\hbar}{2m_ec}$ (and assuming $m_e/m=1$), as a function of the dimensionless ratio $\xi=\hbar (\tilde{\omega} -\omega)/(mc^2)$. This is achieved combining the expressions derived in the text in Eqs.~(\ref{spectrumL1},\ref{spectrumR},\ref{spectrumLp2}).  
We note that at the two critical points ($\xi=\frac{\hbar \omega_\eta}{mc^2}$ and $\xi=\frac{\hbar \omega_\eta}{mc^2}$) the magnetisation presents a discontinuity for each energy level. We remark also that there is a particular value of the magnetic field, which corresponds to the midpoint of the two critical points at which the magnetisation vanishes for each energy level.  }
\end{figure*}

\subsection{Applications to Graphene and Silicene}
Finally we would like to draw attention to  interesting connections of the model presented in this work with the physics of graphene~\cite{Geim:2007aa,Castro-Neto:2009aa} and silicene~\cite{Lalmi:2010kx,Vogt:2012vn,Fleurence:2012ys,Lin:2012zr}, its recently synthetized silicon counterpart. In particular we would like to discuss  the phase transition(s) considered here in the context of these new materials which share several similar properties. In graphene as well as in silicene there are two inequivalent Dirac points, namely the points $K$ and $K'$ of the Brillouin zone. Both experimental~\cite{Chen:2012aa} and theoretical researches~\cite{Ezawa:2013aa} have confirmed that the low energy electronic properties of silicene can be described by a model in which charge carriers  are \emph{massive} Dirac fermions, as opposed to graphene where the charge carriers are massless. The silicene Hamiltonians (in the presence of a homogeneous magnetic field and an oscillator interaction) for the $K$ and $K^\prime$ points are given by~\cite{Ezawa:2012aa,Ezawa:2013aa,Ezawa:2013ab}

\beq
\ba{l}
H_{K^{\phantom{\prime}}}= v_F\,\bm{\sigma^{\phantom{*}}}\cdot\, (\bm{\hat p}-im\omega\beta \bm{\hat x}+\f{e}{c}\bm{{\hat A}})+\b mc^2\,,\\
H_{K^\prime}= v_F\,\bm{\sigma^*}\cdot \, (\bm{\hat p}-im\omega\beta \bm{\hat x}+\f{e}{c}\bm{{\hat A}})+\b mc^2\,,\\
\ea
\eeq
where ${\bm\sigma^*}=(-\sigma_x,\sigma_y)$ and are clearly related to the Hamiltonian of our system of a 2-dimensional Dirac oscillator in constant magnetic field, c.f. Eq.~\eqref{h1}. It is easy to see that one may obtain $H_K$ from Eq.(\ref{h1}) by replacing $c$ by $v_F$ and as a consequence all results, in particular the value of the critical magnetic field in Eq.(\ref{bc}) remain unaltered. Now proceeding in a similar manner it can be shown that for $H_{K^\prime}$ the critical magnetic field ($B_{\text{cr}}'$) is given by
\beq\label{bcs}
B_{\text{cr}}'=-\f{2v_F}{e}\left(m\omega-\f{\eta}{2\hbar}\right)\,.
\eeq
Comparing Eqs.(\ref{bc}) and (\ref{bcs}) we find  the interesting property that the value of the critical magnetic field is different for the two Dirac points. If we have a quantum phase transition at the $K$ point with the critical field in Eq.~\eqref{bc} then there will be no phase transition at the $K'$ point and viceversa if there is a phase transition at the $K'$ point with the critical field in Eq.~\eqref{bcs}, then there will be no phase transition at the $K$ point.  We note that in the absence of non-commutativity in the momentum coordinates ($\eta \to 0$) the critical field at the $K'$ point ($B_{\text{cr}}'$) would be exactly reversed relative to the  critical field of the $K$ point, Eq.~\eqref{bc}. 

Similar considerations apply to the critical field  $B_{\text{cr}}^{*'}$ of the second quantum phase transition at the point $K'$. We find that $B_{\text{cr}}^{*'}$ is given by:
\begin{equation}
\label{bcrit*p}
 B_{\text{cr}}^{*'} = -\frac{2m\, v_F}{e}\left( \omega -\omega_\theta\right)= -\frac{2\,v_F}{e}\left(m\omega - \frac{2\hbar}{\theta}\right)\,,
\end{equation} 
which is to be compared with Eq.~\eqref{bcrit*}. 
If we have a quantum phase transition at the $K'$ point with the critical field in Eq.~\eqref{bcrit*p} then there will be no phase transition at the $K$ point and viceversa if there is a phase transition at the $K$ point with the critical field in Eq.~\eqref{bcrit*}, then there will be no phase transition at the $K'$ point.
Here again, as observed for the second phase transition at the $K$ point, see Eq.~\eqref{bcrit*},  in the absence of non-commutative space coordinates ($\theta \to 0$) $B_{\text{cr}}^{*'}\to \infty$ and there is no  second phase transition.

In the case of graphene the charge carriers are massless and the Hamiltonians for the $K$ and $K^\prime$ points (assuming a Dirac oscillator coupling and an external homogenous magnetic field) are given by \cite{Castro-Neto:2009aa,Quimbay:2013ab}
\beq\label{graphene}
\ba{l}
 H_{K^{\phantom{\prime}}}= v_F\,\bm{\sigma^{\phantom{*}}}\cdot\,(\bm{\hat p}-i\Omega\beta \bm{\hat x}+\f{e}{c}\bm{{\hat A}})\,,\\
H_{K^\prime}= v_F\,\bm{\sigma^*}\cdot\,(\bm{\hat p}-i\Omega\beta \bm{\hat x}+\f{e}{c}\bm{{\hat A}})\,,
\ea
\eeq
where $\bm{\sigma}=(\sigma_x,\sigma_y)$ and $\bm{\sigma^*}=(\sigma_x,-\sigma_y)$. As discussed in \cite{Quimbay:2013ab} the Dirac oscillator coupling in graphene is assumed to arise from an internal effective magnetic field $B_I$ due to the motion of the charge carriers relative to the planar hexagonal arrangement of the carbon atoms~\cite{Quimbay:2013ab}.
Indeed in \cite{Quimbay:2013ab} the oscillator coupling for massless fermions is directly proportional to the effective internal magnetic field $B_I$: $\Omega= e B_I/4$. 
In this case the critical magnetic field  for $H_{K^\prime}$ in the first phase transition is given by: 
\beq
B_{\text{cr}}'=-\frac{2v_F}{e}\left(\Omega-\f{\eta}{2\hbar}\right)\,.
\eeq
Again the critical field at the $K'$ point is different than that at the $K$ point and would be exactly reversed were it not for the non-zero non-commuting parameter $\eta$.
The critical field for the second phase transition at the $K'$ point will now be given by:
\begin{equation}
 B_{\text{cr}}^{*'} =  -\frac{2\,v_F}{e}\left(\Omega - \frac{2\hbar}{\theta}\right)\,.
\end{equation} 
and similar considerations apply as in the massive case of silicene described above.

\section{Conclusions}\label{conclusions}
In the present work we have studied the Dirac oscillator in a constant homogenous magnetic field in the presence of non-commutativity both in the space and momentum coordinates. While the Dirac oscillator has become, over the years, a paradigm of mathematical physics very recently its one dimensional version has been experimentally realized and observed for the first time~\cite{Franco-Villafane:2013aa} with realistic prospects of realizing soon a 2-dimensional version. We were able to solve exactly the corresponding  eigenvalue equations according to the strength of the magnetic field. 
The system in the absence of non-commutativity is 
known to have a left-right chiral phase transition for $B=B_{\text{cr}}=\frac{2mc}{e}\omega$. We find that the non-commutativity of the momentum coordinates ($\eta-$parameter, see Eq.~\eqref{NCEQ}) simply shifts the left-right chiral phase transition. The presence of non-commutativity in the  space coordinates ($\theta-$parameter, see Eq.~\eqref{NCEQ}) changes the picture quite dramatically. It introduces a new right-left phase transition at $B=B_{\text{cr}}^*=\frac{2mc}{e}(\omega+\omega_\theta$) which is absent in the commutative limit (as $\theta\to 0$, $\omega_\theta \to \infty$). 

Our explicit solutions in the three chiral phases (left-right-left) show that for every energy level (with the exception of the singlet-like zero modes) there is a class of states with finite degeneracy and opposite average angular momentum $\langle L_z\rangle $ relative to the class of states with infinite degeneracy. Apparently, these classes of states with finite degeneracy in the left phases and  infinite degeneracy in the right phase, were previously overlooked and the sign of the angular momentum was proposed as an order parameter of the quantum phase transition~\cite{Quimbay:2013aa,Bermudez:2008ab,Bermudez:2007aa,Bermudez:2007ab,Bermudez:2008aa,Mandal:2010aa,Mandal:2012aa}. 

We have discussed the magnetisation of the energy levels of the system and we have shown that this physical observable has the potential of being able to characterise the two quantum phase transitions since it offers a finite discontinuity (for every energy level) at the two critical points. A rather peculiar prediction for this observable is the existence of a \emph{fixed point} with vanishing magnetisation, for every energy level, exactly located at the midpoint of the two critical points.  
Work is in progress to present a full thermodynamic treatment of the system based on the solution presented here.
 
 Finally we have addressed possible implications of our results  in the physics of  silicene, the silicon counterpart of graphene, recently synthesised. In silicene the low energy electron excitations are described by a free massive Dirac equation. A Dirac oscillator coupling could arise  via an effective internal magnetic field generated by the electron motion in the honeycomb lattice. The results presented in this paper could then be directly applied to silicene in an external magnetic field. We have found that the values of the critical fields for which the quantum phase transitions (first and second) take place are different for the $K$ and $K'$ points. While this consideration apply as well for graphene in order to use our detailed explicit solutions for graphene one need to separately consider the massless limit which we demand to a forthcoming study.   

\begin{acknowledgments}
One of us (P.~R.) wishes to thank INFN Sezione di Perugia for supporting a visit during which part of this work was carried out.  He would also like to thank the Physics Department of the University of Perugia for hospitality.
\end{acknowledgments} 

\bibliography{QPT.bib}

\begin{thebibliography}{54}%
\makeatletter
\providecommand \@ifxundefined [1]{%
 \@ifx{#1\undefined}
}%
\providecommand \@ifnum [1]{%
 \ifnum #1\expandafter \@firstoftwo
 \else \expandafter \@secondoftwo
 \fi
}%
\providecommand \@ifx [1]{%
 \ifx #1\expandafter \@firstoftwo
 \else \expandafter \@secondoftwo
 \fi
}%
\providecommand \natexlab [1]{#1}%
\providecommand \enquote  [1]{``#1''}%
\providecommand \bibnamefont  [1]{#1}%
\providecommand \bibfnamefont [1]{#1}%
\providecommand \citenamefont [1]{#1}%
\providecommand \href@noop [0]{\@secondoftwo}%
\providecommand \href [0]{\begingroup \@sanitize@url \@href}%
\providecommand \@href[1]{\@@startlink{#1}\@@href}%
\providecommand \@@href[1]{\endgroup#1\@@endlink}%
\providecommand \@sanitize@url [0]{\catcode `\\12\catcode `\$12\catcode
  `\&12\catcode `\#12\catcode `\^12\catcode `\_12\catcode `\%12\relax}%
\providecommand \@@startlink[1]{}%
\providecommand \@@endlink[0]{}%
\providecommand \url  [0]{\begingroup\@sanitize@url \@url }%
\providecommand \@url [1]{\endgroup\@href {#1}{\urlprefix }}%
\providecommand \urlprefix  [0]{URL }%
\providecommand \Eprint [0]{\href }%
\providecommand \doibase [0]{http://dx.doi.org/}%
\providecommand \selectlanguage [0]{\@gobble}%
\providecommand \bibinfo  [0]{\@secondoftwo}%
\providecommand \bibfield  [0]{\@secondoftwo}%
\providecommand \translation [1]{[#1]}%
\providecommand \BibitemOpen [0]{}%
\providecommand \bibitemStop [0]{}%
\providecommand \bibitemNoStop [0]{.\EOS\space}%
\providecommand \EOS [0]{\spacefactor3000\relax}%
\providecommand \BibitemShut  [1]{\csname bibitem#1\endcsname}%
\let\auto@bib@innerbib\@empty
\bibitem [{\citenamefont {Connes}\ \emph {et~al.}(1998)\citenamefont {Connes},
  \citenamefont {Douglas},\ and\ \citenamefont {Schwarz}}]{Connes:1998aa}%
  \BibitemOpen
  \bibfield  {author} {\bibinfo {author} {\bibfnamefont {A.}~\bibnamefont
  {Connes}}, \bibinfo {author} {\bibfnamefont {M.~R.}\ \bibnamefont {Douglas}},
  \ and\ \bibinfo {author} {\bibfnamefont {A.}~\bibnamefont {Schwarz}},\ }\href
  {http://stacks.iop.org/1126-6708/1998/i=02/a=003} {\bibfield  {journal}
  {\bibinfo  {journal} {J. High Energy Phys.}\ }\textbf {\bibinfo {volume}
  {9802}},\ \bibinfo {pages} {003} (\bibinfo {year} {1998})}\BibitemShut
  {NoStop}%
\bibitem [{\citenamefont {Seiberg}\ and\ \citenamefont
  {Witten}(1999)}]{Seiberg:1999aa}%
  \BibitemOpen
  \bibfield  {author} {\bibinfo {author} {\bibfnamefont {N.}~\bibnamefont
  {Seiberg}}\ and\ \bibinfo {author} {\bibfnamefont {E.}~\bibnamefont
  {Witten}},\ }\href {http://stacks.iop.org/1126-6708/1999/i=09/a=032}
  {\bibfield  {journal} {\bibinfo  {journal} {J. High Energy Phys.}\ }\textbf
  {\bibinfo {volume} {9909}},\ \bibinfo {pages} {032} (\bibinfo {year}
  {1999})}\BibitemShut {NoStop}%
\bibitem [{\citenamefont {Douglas}\ and\ \citenamefont
  {Nekrasov}(2001)}]{Douglas:2001aa}%
  \BibitemOpen
  \bibfield  {author} {\bibinfo {author} {\bibfnamefont {M.~R.}\ \bibnamefont
  {Douglas}}\ and\ \bibinfo {author} {\bibfnamefont {N.~A.}\ \bibnamefont
  {Nekrasov}},\ }\href {\doibase 10.1103/RevModPhys.73.977} {\bibfield
  {journal} {\bibinfo  {journal} {Rev. Mod. Phys.}\ }\textbf {\bibinfo {volume}
  {73}},\ \bibinfo {pages} {977} (\bibinfo {year} {2001})}\BibitemShut
  {NoStop}%
\bibitem [{\citenamefont {Szabo}(2003)}]{Szabo:2003aa}%
  \BibitemOpen
  \bibfield  {author} {\bibinfo {author} {\bibfnamefont {R.~J.}\ \bibnamefont
  {Szabo}},\ }\href {\doibase http://dx.doi.org/10.1016/S0370-1573(03)00059-0}
  {\bibfield  {journal} {\bibinfo  {journal} {Phys. Rept.}\ }\textbf {\bibinfo
  {volume} {378}},\ \bibinfo {pages} {207 } (\bibinfo {year}
  {2003})}\BibitemShut {NoStop}%
\bibitem [{\citenamefont {Duval}\ and\ \citenamefont
  {Horv\'{a}thy}(2000)}]{Duval:2000aa}%
  \BibitemOpen
  \bibfield  {author} {\bibinfo {author} {\bibfnamefont {C.}~\bibnamefont
  {Duval}}\ and\ \bibinfo {author} {\bibfnamefont {P.}~\bibnamefont
  {Horv\'{a}thy}},\ }\href {\doibase
  http://dx.doi.org/10.1016/S0370-2693(00)00341-5} {\bibfield  {journal}
  {\bibinfo  {journal} {Phys. Lett. B}\ }\textbf {\bibinfo {volume} {479}},\
  \bibinfo {pages} {284 } (\bibinfo {year} {2000})}\BibitemShut {NoStop}%
\bibitem [{\citenamefont {Gamboa}\ \emph
  {et~al.}(2001{\natexlab{a}})\citenamefont {Gamboa}, \citenamefont {Loewe},\
  and\ \citenamefont {Rojas}}]{Gamboa:2001aa}%
  \BibitemOpen
  \bibfield  {author} {\bibinfo {author} {\bibfnamefont {J.}~\bibnamefont
  {Gamboa}}, \bibinfo {author} {\bibfnamefont {M.}~\bibnamefont {Loewe}}, \
  and\ \bibinfo {author} {\bibfnamefont {J.~C.}\ \bibnamefont {Rojas}},\ }\href
  {\doibase 10.1103/PhysRevD.64.067901} {\bibfield  {journal} {\bibinfo
  {journal} {Phys. Rev. D}\ }\textbf {\bibinfo {volume} {64}},\ \bibinfo
  {pages} {067901} (\bibinfo {year} {2001}{\natexlab{a}})}\BibitemShut
  {NoStop}%
\bibitem [{\citenamefont {Gamboa}\ \emph
  {et~al.}(2001{\natexlab{b}})\citenamefont {Gamboa}, \citenamefont {Mendez},
  \citenamefont {Loewe},\ and\ \citenamefont {Rojas}}]{Gamboa:2001ab}%
  \BibitemOpen
  \bibfield  {author} {\bibinfo {author} {\bibfnamefont {J.}~\bibnamefont
  {Gamboa}}, \bibinfo {author} {\bibfnamefont {F.}~\bibnamefont {Mendez}},
  \bibinfo {author} {\bibfnamefont {M.}~\bibnamefont {Loewe}}, \ and\ \bibinfo
  {author} {\bibfnamefont {J.~C.}\ \bibnamefont {Rojas}},\ }\href {\doibase
  10.1142/S0217732301005345} {\bibfield  {journal} {\bibinfo  {journal} {Mod.
  Phys. Lett. A}\ }\textbf {\bibinfo {volume} {16}},\ \bibinfo {pages} {2075}
  (\bibinfo {year} {2001}{\natexlab{b}})}\BibitemShut {NoStop}%
\bibitem [{\citenamefont {Nair}\ and\ \citenamefont
  {Polychronakos}(2001)}]{Nair:2001aa}%
  \BibitemOpen
  \bibfield  {author} {\bibinfo {author} {\bibfnamefont {V.}~\bibnamefont
  {Nair}}\ and\ \bibinfo {author} {\bibfnamefont {A.}~\bibnamefont
  {Polychronakos}},\ }\href {\doibase
  http://dx.doi.org/10.1016/S0370-2693(01)00339-2} {\bibfield  {journal}
  {\bibinfo  {journal} {Phys. Lett. B}\ }\textbf {\bibinfo {volume} {505}},\
  \bibinfo {pages} {267 } (\bibinfo {year} {2001})}\BibitemShut {NoStop}%
\bibitem [{\citenamefont {Gamboa}\ \emph {et~al.}(2002)\citenamefont {Gamboa},
  \citenamefont {M\'{e}ndez}, \citenamefont {Loewe},\ and\ \citenamefont
  {Rojas}}]{Gamboa:2002aa}%
  \BibitemOpen
  \bibfield  {author} {\bibinfo {author} {\bibfnamefont {J.}~\bibnamefont
  {Gamboa}}, \bibinfo {author} {\bibfnamefont {F.}~\bibnamefont {M\'{e}ndez}},
  \bibinfo {author} {\bibfnamefont {M.}~\bibnamefont {Loewe}}, \ and\ \bibinfo
  {author} {\bibfnamefont {J.~C.}\ \bibnamefont {Rojas}},\ }\href {\doibase
  10.1142/S0217751X02010960} {\bibfield  {journal} {\bibinfo  {journal} {Int.
  J. Mod. Phys. A}\ }\textbf {\bibinfo {volume} {17}},\ \bibinfo {pages} {2555}
  (\bibinfo {year} {2002})}\BibitemShut {NoStop}%
\bibitem [{\citenamefont {Smailagic}\ and\ \citenamefont
  {Spallucci}(2002{\natexlab{a}})}]{Smailagic:2002aa}%
  \BibitemOpen
  \bibfield  {author} {\bibinfo {author} {\bibfnamefont {A.}~\bibnamefont
  {Smailagic}}\ and\ \bibinfo {author} {\bibfnamefont {E.}~\bibnamefont
  {Spallucci}},\ }\href {\doibase 10.1103/PhysRevD.65.107701} {\bibfield
  {journal} {\bibinfo  {journal} {Phys. Rev. D}\ }\textbf {\bibinfo {volume}
  {65}},\ \bibinfo {pages} {107701} (\bibinfo {year}
  {2002}{\natexlab{a}})}\BibitemShut {NoStop}%
\bibitem [{\citenamefont {Smailagic}\ and\ \citenamefont
  {Spallucci}(2002{\natexlab{b}})}]{Smailagic:2002ab}%
  \BibitemOpen
  \bibfield  {author} {\bibinfo {author} {\bibfnamefont {A.}~\bibnamefont
  {Smailagic}}\ and\ \bibinfo {author} {\bibfnamefont {E.}~\bibnamefont
  {Spallucci}},\ }\href {http://stacks.iop.org/0305-4470/35/i=26/a=103}
  {\bibfield  {journal} {\bibinfo  {journal} {J. Phys. A: Math. Gen.}\ }\textbf
  {\bibinfo {volume} {35}},\ \bibinfo {pages} {L363} (\bibinfo {year}
  {2002}{\natexlab{b}})}\BibitemShut {NoStop}%
\bibitem [{\citenamefont {Bellucci}\ \emph {et~al.}(2001)\citenamefont
  {Bellucci}, \citenamefont {Nersessian},\ and\ \citenamefont
  {Sochichiu}}]{Bellucci:2001aa}%
  \BibitemOpen
  \bibfield  {author} {\bibinfo {author} {\bibfnamefont {S.}~\bibnamefont
  {Bellucci}}, \bibinfo {author} {\bibfnamefont {A.}~\bibnamefont
  {Nersessian}}, \ and\ \bibinfo {author} {\bibfnamefont {C.}~\bibnamefont
  {Sochichiu}},\ }\href {\doibase
  http://dx.doi.org/10.1016/S0370-2693(01)01304-1} {\bibfield  {journal}
  {\bibinfo  {journal} {Phys. Lett. B}\ }\textbf {\bibinfo {volume} {522}},\
  \bibinfo {pages} {345 } (\bibinfo {year} {2001})}\BibitemShut {NoStop}%
\bibitem [{\citenamefont {Bertolami}\ \emph {et~al.}(2005)\citenamefont
  {Bertolami}, \citenamefont {Rosa}, \citenamefont {de~Arag\~ao}, \citenamefont
  {Castorina},\ and\ \citenamefont {Zappal\`a}}]{Bertolami:2005aa}%
  \BibitemOpen
  \bibfield  {author} {\bibinfo {author} {\bibfnamefont {O.}~\bibnamefont
  {Bertolami}}, \bibinfo {author} {\bibfnamefont {J.~G.}\ \bibnamefont {Rosa}},
  \bibinfo {author} {\bibfnamefont {C.~M.~L.}\ \bibnamefont {de~Arag\~ao}},
  \bibinfo {author} {\bibfnamefont {P.}~\bibnamefont {Castorina}}, \ and\
  \bibinfo {author} {\bibfnamefont {D.}~\bibnamefont {Zappal\`a}},\ }\href
  {\doibase 10.1103/PhysRevD.72.025010} {\bibfield  {journal} {\bibinfo
  {journal} {Phys. Rev. D}\ }\textbf {\bibinfo {volume} {72}},\ \bibinfo
  {pages} {025010} (\bibinfo {year} {2005})}\BibitemShut {NoStop}%
\bibitem [{\citenamefont {Horv\'{a}thy}\ \emph {et~al.}(2010)\citenamefont
  {Horv\'{a}thy}, \citenamefont {Martina},\ and\ \citenamefont
  {Stichel}}]{Horvathy:2010wv}%
  \BibitemOpen
  \bibfield  {author} {\bibinfo {author} {\bibfnamefont {P.}~\bibnamefont
  {Horv\'{a}thy}}, \bibinfo {author} {\bibfnamefont {L.}~\bibnamefont
  {Martina}}, \ and\ \bibinfo {author} {\bibfnamefont {P.}~\bibnamefont
  {Stichel}},\ }\href {\doibase 10.3842/SIGMA.2010.060} {\bibfield  {journal}
  {\bibinfo  {journal} {SIGMA}\ }\textbf {\bibinfo {volume} {6}},\ \bibinfo
  {pages} {060} (\bibinfo {year} {2010})},\ \Eprint
  {http://arxiv.org/abs/1002.4772} {arXiv:1002.4772 [hep-th]} \BibitemShut
  {NoStop}%
\bibitem [{\citenamefont {Ito}\ \emph {et~al.}(1967)\citenamefont {Ito},
  \citenamefont {Mori},\ and\ \citenamefont {Carriere}}]{Ito:1967aa}%
  \BibitemOpen
  \bibfield  {author} {\bibinfo {author} {\bibfnamefont {D.}~\bibnamefont
  {Ito}}, \bibinfo {author} {\bibfnamefont {K.}~\bibnamefont {Mori}}, \ and\
  \bibinfo {author} {\bibfnamefont {E.}~\bibnamefont {Carriere}},\ }\href@noop
  {} {\bibfield  {journal} {\bibinfo  {journal} {Nuovo Cimento A}\ }\textbf
  {\bibinfo {volume} {51}},\ \bibinfo {pages} {1119} (\bibinfo {year}
  {1967})}\BibitemShut {NoStop}%
\bibitem [{\citenamefont {Cook}(1971)}]{Cook:1971aa}%
  \BibitemOpen
  \bibfield  {author} {\bibinfo {author} {\bibfnamefont {P.}~\bibnamefont
  {Cook}},\ }\href {\doibase 10.1007/BF02785170} {\bibfield  {journal}
  {\bibinfo  {journal} {Lett. Nuovo Cimento}\ }\textbf {\bibinfo {volume}
  {1}},\ \bibinfo {pages} {419} (\bibinfo {year} {1971})}\BibitemShut {NoStop}%
\bibitem [{\citenamefont {Moshinsky}\ and\ \citenamefont
  {Szczepaniak}(1989)}]{Moshinsky:1989aa}%
  \BibitemOpen
  \bibfield  {author} {\bibinfo {author} {\bibfnamefont {M.}~\bibnamefont
  {Moshinsky}}\ and\ \bibinfo {author} {\bibfnamefont {A.}~\bibnamefont
  {Szczepaniak}},\ }\href {http://stacks.iop.org/0305-4470/22/i=17/a=002}
  {\bibfield  {journal} {\bibinfo  {journal} {J. Phys. A: Math. Gen.}\ }\textbf
  {\bibinfo {volume} {22}},\ \bibinfo {pages} {L817} (\bibinfo {year}
  {1989})}\BibitemShut {NoStop}%
\bibitem [{\citenamefont {Benitez}\ \emph {et~al.}(1990)\citenamefont
  {Benitez}, \citenamefont {Mart{i}nez~y Romero}, \citenamefont
  {Nu{n}ez-Y{e}pez},\ and\ \citenamefont {Salas-Brito}}]{Benitez:1990aa}%
  \BibitemOpen
  \bibfield  {author} {\bibinfo {author} {\bibfnamefont {J.}~\bibnamefont
  {Benitez}}, \bibinfo {author} {\bibfnamefont {R.~P.}\ \bibnamefont
  {Mart{i}nez~y Romero}}, \bibinfo {author} {\bibfnamefont {H.~N.}\
  \bibnamefont {Nu{n}ez-Y{e}pez}}, \ and\ \bibinfo {author} {\bibfnamefont
  {A.~L.}\ \bibnamefont {Salas-Brito}},\ }\href {\doibase
  10.1103/PhysRevLett.64.1643} {\bibfield  {journal} {\bibinfo  {journal}
  {Phys. Rev. Lett.}\ }\textbf {\bibinfo {volume} {64}},\ \bibinfo {pages}
  {1643} (\bibinfo {year} {1990})}\BibitemShut {NoStop}%
\bibitem [{\citenamefont {Rozmej}\ and\ \citenamefont
  {Arvieu}(1999)}]{Rozmej:1999aa}%
  \BibitemOpen
  \bibfield  {author} {\bibinfo {author} {\bibfnamefont {P.}~\bibnamefont
  {Rozmej}}\ and\ \bibinfo {author} {\bibfnamefont {R.}~\bibnamefont
  {Arvieu}},\ }\href {http://stacks.iop.org/0305-4470/32/i=28/a=314} {\bibfield
   {journal} {\bibinfo  {journal} {J. Phys. A: Math. Gen.}\ }\textbf {\bibinfo
  {volume} {32}},\ \bibinfo {pages} {5367} (\bibinfo {year}
  {1999})}\BibitemShut {NoStop}%
\bibitem [{\citenamefont {Mirza}\ and\ \citenamefont
  {Mohadesi}(2004)}]{Mirza:2004aa}%
  \BibitemOpen
  \bibfield  {author} {\bibinfo {author} {\bibfnamefont {B.}~\bibnamefont
  {Mirza}}\ and\ \bibinfo {author} {\bibfnamefont {M.}~\bibnamefont
  {Mohadesi}},\ }\href {http://stacks.iop.org/0253-6102/42/i=5/a=664}
  {\bibfield  {journal} {\bibinfo  {journal} {Comm. Th. Phys.}\ }\textbf
  {\bibinfo {volume} {42}},\ \bibinfo {pages} {664} (\bibinfo {year}
  {2004})}\BibitemShut {NoStop}%
\bibitem [{\citenamefont {Sadurni}\ \emph
  {et~al.}(2010{\natexlab{a}})\citenamefont {Sadurni}, \citenamefont {Torres},\
  and\ \citenamefont {Seligman}}]{Sadurni:2010aa}%
  \BibitemOpen
  \bibfield  {author} {\bibinfo {author} {\bibfnamefont {E.}~\bibnamefont
  {Sadurni}}, \bibinfo {author} {\bibfnamefont {J.~M.}\ \bibnamefont {Torres}},
  \ and\ \bibinfo {author} {\bibfnamefont {T.~H.}\ \bibnamefont {Seligman}},\
  }\href {http://stacks.iop.org/1751-8121/43/i=28/a=285204} {\bibfield
  {journal} {\bibinfo  {journal} {J. Phys. A: Math. Gen.}\ }\textbf {\bibinfo
  {volume} {43}},\ \bibinfo {pages} {285204} (\bibinfo {year}
  {2010}{\natexlab{a}})}\BibitemShut {NoStop}%
\bibitem [{\citenamefont {Boumali}\ and\ \citenamefont
  {Hassanabadi}(2013)}]{Boumali:2013aa}%
  \BibitemOpen
  \bibfield  {author} {\bibinfo {author} {\bibfnamefont {A.}~\bibnamefont
  {Boumali}}\ and\ \bibinfo {author} {\bibfnamefont {H.}~\bibnamefont
  {Hassanabadi}},\ }\href {\doibase 10.1140/epjp/i2013-13124-y} {\bibfield
  {journal} {\bibinfo  {journal} {Eur. Phys. J. Plus}\ }\textbf {\bibinfo
  {volume} {128}},\ \bibinfo {eid} {124} (\bibinfo {year} {2013}),\
  10.1140/epjp/i2013-13124-y}\BibitemShut {NoStop}%
\bibitem [{\citenamefont {Faessler}\ \emph {et~al.}(2005)\citenamefont
  {Faessler}, \citenamefont {Kukulin},\ and\ \citenamefont
  {Shikhalev}}]{Faessler:2005aa}%
  \BibitemOpen
  \bibfield  {author} {\bibinfo {author} {\bibfnamefont {A.}~\bibnamefont
  {Faessler}}, \bibinfo {author} {\bibfnamefont {V.}~\bibnamefont {Kukulin}}, \
  and\ \bibinfo {author} {\bibfnamefont {M.}~\bibnamefont {Shikhalev}},\ }\href
  {\doibase http://dx.doi.org/10.1016/j.aop.2005.05.008} {\bibfield  {journal}
  {\bibinfo  {journal} {Annals Phys.}\ }\textbf {\bibinfo {volume} {320}},\
  \bibinfo {pages} {71 } (\bibinfo {year} {2005})}\BibitemShut {NoStop}%
\bibitem [{\citenamefont {Grineviciute}\ and\ \citenamefont
  {Halderson}(2012)}]{Grineviciute:2012aa}%
  \BibitemOpen
  \bibfield  {author} {\bibinfo {author} {\bibfnamefont {J.}~\bibnamefont
  {Grineviciute}}\ and\ \bibinfo {author} {\bibfnamefont {D.}~\bibnamefont
  {Halderson}},\ }\href {\doibase 10.1103/PhysRevC.85.054617} {\bibfield
  {journal} {\bibinfo  {journal} {Phys. Rev. C}\ }\textbf {\bibinfo {volume}
  {85}},\ \bibinfo {pages} {054617} (\bibinfo {year} {2012})}\BibitemShut
  {NoStop}%
\bibitem [{\citenamefont {Munarriz}\ \emph {et~al.}(2012)\citenamefont
  {Munarriz}, \citenamefont {Dominguez-Adame},\ and\ \citenamefont
  {Lima}}]{Munarriz:2012aa}%
  \BibitemOpen
  \bibfield  {author} {\bibinfo {author} {\bibfnamefont {J.}~\bibnamefont
  {Munarriz}}, \bibinfo {author} {\bibfnamefont {F.}~\bibnamefont
  {Dominguez-Adame}}, \ and\ \bibinfo {author} {\bibfnamefont {R.}~\bibnamefont
  {Lima}},\ }\href {\doibase http://dx.doi.org/10.1016/j.physleta.2012.10.029}
  {\bibfield  {journal} {\bibinfo  {journal} {Phys. Lett. A}\ }\textbf
  {\bibinfo {volume} {376}},\ \bibinfo {pages} {3475 } (\bibinfo {year}
  {2012})}\BibitemShut {NoStop}%
\bibitem [{\citenamefont {Romera}(2011)}]{Romera:2011aa}%
  \BibitemOpen
  \bibfield  {author} {\bibinfo {author} {\bibfnamefont {E.}~\bibnamefont
  {Romera}},\ }\href {\doibase 10.1103/PhysRevA.84.052102} {\bibfield
  {journal} {\bibinfo  {journal} {Phys. Rev. A}\ }\textbf {\bibinfo {volume}
  {84}},\ \bibinfo {pages} {052102} (\bibinfo {year} {2011})}\BibitemShut
  {NoStop}%
\bibitem [{\citenamefont {Wang}\ \emph {et~al.}(2012)\citenamefont {Wang},
  \citenamefont {Cao},\ and\ \citenamefont {Xiong}}]{Wang:2012aa}%
  \BibitemOpen
  \bibfield  {author} {\bibinfo {author} {\bibfnamefont {Y.}~\bibnamefont
  {Wang}}, \bibinfo {author} {\bibfnamefont {J.}~\bibnamefont {Cao}}, \ and\
  \bibinfo {author} {\bibfnamefont {S.}~\bibnamefont {Xiong}},\ }\href
  {\doibase 10.1140/epjb/e2012-30243-7} {\bibfield  {journal} {\bibinfo
  {journal} {Eur. Phys. J. B}\ }\textbf {\bibinfo {volume} {85}},\ \bibinfo
  {eid} {237} (\bibinfo {year} {2012}),\
  10.1140/epjb/e2012-30243-7}\BibitemShut {NoStop}%
\bibitem [{\citenamefont {Bermudez}\ \emph
  {et~al.}(2007{\natexlab{a}})\citenamefont {Bermudez}, \citenamefont
  {Martin-Delgado},\ and\ \citenamefont {Solano}}]{Bermudez:2007ab}%
  \BibitemOpen
  \bibfield  {author} {\bibinfo {author} {\bibfnamefont {A.}~\bibnamefont
  {Bermudez}}, \bibinfo {author} {\bibfnamefont {M.~A.}\ \bibnamefont
  {Martin-Delgado}}, \ and\ \bibinfo {author} {\bibfnamefont {E.}~\bibnamefont
  {Solano}},\ }\href {\doibase 10.1103/PhysRevLett.99.123602} {\bibfield
  {journal} {\bibinfo  {journal} {Phys. Rev. Lett.}\ }\textbf {\bibinfo
  {volume} {99}},\ \bibinfo {pages} {123602} (\bibinfo {year}
  {2007}{\natexlab{a}})}\BibitemShut {NoStop}%
\bibitem [{\citenamefont {Dodonov}(2002)}]{Dodonov:2002aa}%
  \BibitemOpen
  \bibfield  {author} {\bibinfo {author} {\bibfnamefont {V.~V.}\ \bibnamefont
  {Dodonov}},\ }\href {http://stacks.iop.org/1464-4266/4/i=1/a=201} {\bibfield
  {journal} {\bibinfo  {journal} {J. Opt. B: Quantum S. O.}\ }\textbf {\bibinfo
  {volume} {4}},\ \bibinfo {pages} {R1} (\bibinfo {year} {2002})}\BibitemShut
  {NoStop}%
\bibitem [{\citenamefont {Lamata}\ \emph {et~al.}(2007)\citenamefont {Lamata},
  \citenamefont {Le\'on}, \citenamefont {Sch\"atz},\ and\ \citenamefont
  {Solano}}]{Lamata:2007aa}%
  \BibitemOpen
  \bibfield  {author} {\bibinfo {author} {\bibfnamefont {L.}~\bibnamefont
  {Lamata}}, \bibinfo {author} {\bibfnamefont {J.}~\bibnamefont {Le\'on}},
  \bibinfo {author} {\bibfnamefont {T.}~\bibnamefont {Sch\"atz}}, \ and\
  \bibinfo {author} {\bibfnamefont {E.}~\bibnamefont {Solano}},\ }\href
  {\doibase 10.1103/PhysRevLett.98.253005} {\bibfield  {journal} {\bibinfo
  {journal} {Phys. Rev. Lett.}\ }\textbf {\bibinfo {volume} {98}},\ \bibinfo
  {pages} {253005} (\bibinfo {year} {2007})}\BibitemShut {NoStop}%
\bibitem [{\citenamefont {Franco-Villafane}\ \emph {et~al.}(2013)\citenamefont
  {Franco-Villafane}, \citenamefont {Sadurni}, \citenamefont {Barkhofen},
  \citenamefont {Kuhl}, \citenamefont {Mortessagne},\ and\ \citenamefont
  {Seligman}}]{Franco-Villafane:2013aa}%
  \BibitemOpen
  \bibfield  {author} {\bibinfo {author} {\bibfnamefont {J.~A.}\ \bibnamefont
  {Franco-Villafane}}, \bibinfo {author} {\bibfnamefont {E.}~\bibnamefont
  {Sadurni}}, \bibinfo {author} {\bibfnamefont {S.}~\bibnamefont {Barkhofen}},
  \bibinfo {author} {\bibfnamefont {U.}~\bibnamefont {Kuhl}}, \bibinfo {author}
  {\bibfnamefont {F.}~\bibnamefont {Mortessagne}}, \ and\ \bibinfo {author}
  {\bibfnamefont {T.~H.}\ \bibnamefont {Seligman}},\ }\href {\doibase
  10.1103/PhysRevLett.111.170405} {\bibfield  {journal} {\bibinfo  {journal}
  {Phys. Rev. Lett.}\ }\textbf {\bibinfo {volume} {111}},\ \bibinfo {pages}
  {170405} (\bibinfo {year} {2013})}\BibitemShut {NoStop}%
\bibitem [{\citenamefont {Hul}\ \emph {et~al.}(2004)\citenamefont {Hul},
  \citenamefont {Bauch}, \citenamefont {Pako\ifmmode~\acute{n}\else
  \'{n}\fi{}ski}, \citenamefont {Savytskyy}, \citenamefont
  {\ifmmode~\dot{Z}\else \.{Z}\fi{}yczkowski},\ and\ \citenamefont
  {Sirko}}]{Hul:2004aa}%
  \BibitemOpen
  \bibfield  {author} {\bibinfo {author} {\bibfnamefont {O.}~\bibnamefont
  {Hul}}, \bibinfo {author} {\bibfnamefont {S.}~\bibnamefont {Bauch}}, \bibinfo
  {author} {\bibfnamefont {P.}~\bibnamefont {Pako\ifmmode~\acute{n}\else
  \'{n}\fi{}ski}}, \bibinfo {author} {\bibfnamefont {N.}~\bibnamefont
  {Savytskyy}}, \bibinfo {author} {\bibfnamefont {K.}~\bibnamefont
  {\ifmmode~\dot{Z}\else \.{Z}\fi{}yczkowski}}, \ and\ \bibinfo {author}
  {\bibfnamefont {L.}~\bibnamefont {Sirko}},\ }\href {\doibase
  10.1103/PhysRevE.69.056205} {\bibfield  {journal} {\bibinfo  {journal} {Phys.
  Rev. E}\ }\textbf {\bibinfo {volume} {69}},\ \bibinfo {pages} {056205}
  (\bibinfo {year} {2004})}\BibitemShut {NoStop}%
\bibitem [{\citenamefont {Sadurni}\ \emph
  {et~al.}(2010{\natexlab{b}})\citenamefont {Sadurni}, \citenamefont
  {Seligman},\ and\ \citenamefont {Mortessagne}}]{Sadurni:2010fk}%
  \BibitemOpen
  \bibfield  {author} {\bibinfo {author} {\bibfnamefont {E.}~\bibnamefont
  {Sadurni}}, \bibinfo {author} {\bibfnamefont {T.~H.}\ \bibnamefont
  {Seligman}}, \ and\ \bibinfo {author} {\bibfnamefont {F.}~\bibnamefont
  {Mortessagne}},\ }\href {http://stacks.iop.org/1367-2630/12/i=5/a=053014}
  {\bibfield  {journal} {\bibinfo  {journal} {New J. Phys.}\ }\textbf {\bibinfo
  {volume} {12}},\ \bibinfo {pages} {053014} (\bibinfo {year}
  {2010}{\natexlab{b}})}\BibitemShut {NoStop}%
\bibitem [{\citenamefont {Hul}\ \emph {et~al.}(2005)\citenamefont {Hul},
  \citenamefont {Tymoshchuk}, \citenamefont {Bauch}, \citenamefont {Koch},\
  and\ \citenamefont {Sirko}}]{Hul:2005uq}%
  \BibitemOpen
  \bibfield  {author} {\bibinfo {author} {\bibfnamefont {O.}~\bibnamefont
  {Hul}}, \bibinfo {author} {\bibfnamefont {O.}~\bibnamefont {Tymoshchuk}},
  \bibinfo {author} {\bibfnamefont {S.}~\bibnamefont {Bauch}}, \bibinfo
  {author} {\bibfnamefont {P.~M.}\ \bibnamefont {Koch}}, \ and\ \bibinfo
  {author} {\bibfnamefont {L.}~\bibnamefont {Sirko}},\ }\href
  {http://stacks.iop.org/0305-4470/38/i=49/a=003} {\bibfield  {journal}
  {\bibinfo  {journal} {J. Phys. A: Math. Gen.}\ }\textbf {\bibinfo {volume}
  {38}},\ \bibinfo {pages} {10489} (\bibinfo {year} {2005})}\BibitemShut
  {NoStop}%
\bibitem [{\citenamefont {Bermudez}\ \emph
  {et~al.}(2007{\natexlab{b}})\citenamefont {Bermudez}, \citenamefont
  {Martin-Delgado},\ and\ \citenamefont {Solano}}]{Bermudez:2007aa}%
  \BibitemOpen
  \bibfield  {author} {\bibinfo {author} {\bibfnamefont {A.}~\bibnamefont
  {Bermudez}}, \bibinfo {author} {\bibfnamefont {M.~A.}\ \bibnamefont
  {Martin-Delgado}}, \ and\ \bibinfo {author} {\bibfnamefont {E.}~\bibnamefont
  {Solano}},\ }\href {\doibase 10.1103/PhysRevA.76.041801} {\bibfield
  {journal} {\bibinfo  {journal} {Phys. Rev. A}\ }\textbf {\bibinfo {volume}
  {76}},\ \bibinfo {pages} {041801} (\bibinfo {year}
  {2007}{\natexlab{b}})}\BibitemShut {NoStop}%
\bibitem [{\citenamefont {Bermudez}\ \emph
  {et~al.}(2008{\natexlab{a}})\citenamefont {Bermudez}, \citenamefont
  {Martin-Delgado},\ and\ \citenamefont {Luis}}]{Bermudez:2008aa}%
  \BibitemOpen
  \bibfield  {author} {\bibinfo {author} {\bibfnamefont {A.}~\bibnamefont
  {Bermudez}}, \bibinfo {author} {\bibfnamefont {M.~A.}\ \bibnamefont
  {Martin-Delgado}}, \ and\ \bibinfo {author} {\bibfnamefont {A.}~\bibnamefont
  {Luis}},\ }\href {\doibase 10.1103/PhysRevA.77.033832} {\bibfield  {journal}
  {\bibinfo  {journal} {Phys. Rev. A}\ }\textbf {\bibinfo {volume} {77}},\
  \bibinfo {pages} {033832} (\bibinfo {year} {2008}{\natexlab{a}})}\BibitemShut
  {NoStop}%
\bibitem [{\citenamefont {Mandal}\ and\ \citenamefont
  {Verma}(2010)}]{Mandal:2010aa}%
  \BibitemOpen
  \bibfield  {author} {\bibinfo {author} {\bibfnamefont {B.~P.}\ \bibnamefont
  {Mandal}}\ and\ \bibinfo {author} {\bibfnamefont {S.}~\bibnamefont {Verma}},\
  }\href {\doibase http://dx.doi.org/10.1016/j.physleta.2009.12.048} {\bibfield
   {journal} {\bibinfo  {journal} {Phys. Lett. A}\ }\textbf {\bibinfo {volume}
  {374}},\ \bibinfo {pages} {1021 } (\bibinfo {year} {2010})}\BibitemShut
  {NoStop}%
\bibitem [{\citenamefont {Mandal}\ and\ \citenamefont
  {Rai}(2012)}]{Mandal:2012aa}%
  \BibitemOpen
  \bibfield  {author} {\bibinfo {author} {\bibfnamefont {B.~P.}\ \bibnamefont
  {Mandal}}\ and\ \bibinfo {author} {\bibfnamefont {S.~K.}\ \bibnamefont
  {Rai}},\ }\href {\doibase http://dx.doi.org/10.1016/j.physleta.2012.07.001}
  {\bibfield  {journal} {\bibinfo  {journal} {Phys. Lett. A}\ }\textbf
  {\bibinfo {volume} {376}},\ \bibinfo {pages} {2467 } (\bibinfo {year}
  {2012})}\BibitemShut {NoStop}%
\bibitem [{\citenamefont {Bermudez}\ \emph
  {et~al.}(2008{\natexlab{b}})\citenamefont {Bermudez}, \citenamefont
  {Martin-Delgado},\ and\ \citenamefont {Luis}}]{Bermudez:2008ab}%
  \BibitemOpen
  \bibfield  {author} {\bibinfo {author} {\bibfnamefont {A.}~\bibnamefont
  {Bermudez}}, \bibinfo {author} {\bibfnamefont {M.~A.}\ \bibnamefont
  {Martin-Delgado}}, \ and\ \bibinfo {author} {\bibfnamefont {A.}~\bibnamefont
  {Luis}},\ }\href {\doibase 10.1103/PhysRevA.77.063815} {\bibfield  {journal}
  {\bibinfo  {journal} {Phys. Rev. A}\ }\textbf {\bibinfo {volume} {77}},\
  \bibinfo {pages} {063815} (\bibinfo {year} {2008}{\natexlab{b}})}\BibitemShut
  {NoStop}%
\bibitem [{\citenamefont {Quimbay}\ and\ \citenamefont
  {Strange}(2013{\natexlab{a}})}]{Quimbay:2013aa}%
  \BibitemOpen
  \bibfield  {author} {\bibinfo {author} {\bibfnamefont {C.}~\bibnamefont
  {Quimbay}}\ and\ \bibinfo {author} {\bibfnamefont {P.}~\bibnamefont
  {Strange}},\ }\href@noop {} {\  (\bibinfo {year} {2013}{\natexlab{a}})},\
  \Eprint {http://arxiv.org/abs/1312.5251} {arXiv:1312.5251 [quant-ph]}
  \BibitemShut {NoStop}%
\bibitem [{\citenamefont {Geim}\ and\ \citenamefont
  {Novoselov}(2007)}]{Geim:2007aa}%
  \BibitemOpen
  \bibfield  {author} {\bibinfo {author} {\bibfnamefont {A.~K.}\ \bibnamefont
  {Geim}}\ and\ \bibinfo {author} {\bibfnamefont {K.~S.}\ \bibnamefont
  {Novoselov}},\ }\href {http://dx.doi.org/10.1038/nmat1849} {\bibfield
  {journal} {\bibinfo  {journal} {Nat Mater}\ }\textbf {\bibinfo {volume}
  {6}},\ \bibinfo {pages} {183} (\bibinfo {year} {2007})}\BibitemShut {NoStop}%
\bibitem [{\citenamefont {Castro~Neto}\ \emph {et~al.}(2009)\citenamefont
  {Castro~Neto}, \citenamefont {Guinea}, \citenamefont {Peres}, \citenamefont
  {Novoselov},\ and\ \citenamefont {Geim}}]{Castro-Neto:2009aa}%
  \BibitemOpen
  \bibfield  {author} {\bibinfo {author} {\bibfnamefont {A.~H.}\ \bibnamefont
  {Castro~Neto}}, \bibinfo {author} {\bibfnamefont {F.}~\bibnamefont {Guinea}},
  \bibinfo {author} {\bibfnamefont {N.~M.~R.}\ \bibnamefont {Peres}}, \bibinfo
  {author} {\bibfnamefont {K.~S.}\ \bibnamefont {Novoselov}}, \ and\ \bibinfo
  {author} {\bibfnamefont {A.~K.}\ \bibnamefont {Geim}},\ }\href {\doibase
  10.1103/RevModPhys.81.109} {\bibfield  {journal} {\bibinfo  {journal} {Rev.
  Mod. Phys.}\ }\textbf {\bibinfo {volume} {81}},\ \bibinfo {pages} {109}
  (\bibinfo {year} {2009})}\BibitemShut {NoStop}%
\bibitem [{\citenamefont {Semenoff}(1984)}]{Semenoff:1984aa}%
  \BibitemOpen
  \bibfield  {author} {\bibinfo {author} {\bibfnamefont {G.~W.}\ \bibnamefont
  {Semenoff}},\ }\href {\doibase 10.1103/PhysRevLett.53.2449} {\bibfield
  {journal} {\bibinfo  {journal} {Phys. Rev. Lett.}\ }\textbf {\bibinfo
  {volume} {53}},\ \bibinfo {pages} {2449} (\bibinfo {year}
  {1984})}\BibitemShut {NoStop}%
\bibitem [{\citenamefont {Lalmi}\ \emph {et~al.}(2010)\citenamefont {Lalmi},
  \citenamefont {Oughaddou}, \citenamefont {Enriquez}, \citenamefont {Kara},
  \citenamefont {Vizzini}, \citenamefont {Ealet},\ and\ \citenamefont
  {Aufray}}]{Lalmi:2010kx}%
  \BibitemOpen
  \bibfield  {author} {\bibinfo {author} {\bibfnamefont {B.}~\bibnamefont
  {Lalmi}}, \bibinfo {author} {\bibfnamefont {H.}~\bibnamefont {Oughaddou}},
  \bibinfo {author} {\bibfnamefont {H.}~\bibnamefont {Enriquez}}, \bibinfo
  {author} {\bibfnamefont {A.}~\bibnamefont {Kara}}, \bibinfo {author}
  {\bibfnamefont {S.}~\bibnamefont {Vizzini}}, \bibinfo {author} {\bibfnamefont
  {B.}~\bibnamefont {Ealet}}, \ and\ \bibinfo {author} {\bibfnamefont
  {B.}~\bibnamefont {Aufray}},\ }\href {\doibase
  http://dx.doi.org/10.1063/1.3524215} {\bibfield  {journal} {\bibinfo
  {journal} {Appl. Phys. Lett.}\ }\textbf {\bibinfo {volume} {97}},\ \bibinfo
  {eid} {223109} (\bibinfo {year} {2010})}\BibitemShut {NoStop}%
\bibitem [{\citenamefont {Vogt}\ \emph {et~al.}(2012)\citenamefont {Vogt},
  \citenamefont {De~Padova}, \citenamefont {Quaresima}, \citenamefont {Avila},
  \citenamefont {Frantzeskakis}, \citenamefont {Asensio}, \citenamefont
  {Resta}, \citenamefont {Ealet},\ and\ \citenamefont {Le~Lay}}]{Vogt:2012vn}%
  \BibitemOpen
  \bibfield  {author} {\bibinfo {author} {\bibfnamefont {P.}~\bibnamefont
  {Vogt}}, \bibinfo {author} {\bibfnamefont {P.}~\bibnamefont {De~Padova}},
  \bibinfo {author} {\bibfnamefont {C.}~\bibnamefont {Quaresima}}, \bibinfo
  {author} {\bibfnamefont {J.}~\bibnamefont {Avila}}, \bibinfo {author}
  {\bibfnamefont {E.}~\bibnamefont {Frantzeskakis}}, \bibinfo {author}
  {\bibfnamefont {M.~C.}\ \bibnamefont {Asensio}}, \bibinfo {author}
  {\bibfnamefont {A.}~\bibnamefont {Resta}}, \bibinfo {author} {\bibfnamefont
  {B.}~\bibnamefont {Ealet}}, \ and\ \bibinfo {author} {\bibfnamefont
  {G.}~\bibnamefont {Le~Lay}},\ }\href {\doibase
  10.1103/PhysRevLett.108.155501} {\bibfield  {journal} {\bibinfo  {journal}
  {Phys. Rev. Lett.}\ }\textbf {\bibinfo {volume} {108}},\ \bibinfo {pages}
  {155501} (\bibinfo {year} {2012})}\BibitemShut {NoStop}%
\bibitem [{\citenamefont {Fleurence}\ \emph {et~al.}(2012)\citenamefont
  {Fleurence}, \citenamefont {Friedlein}, \citenamefont {Ozaki}, \citenamefont
  {Kawai}, \citenamefont {Wang},\ and\ \citenamefont
  {Yamada-Takamura}}]{Fleurence:2012ys}%
  \BibitemOpen
  \bibfield  {author} {\bibinfo {author} {\bibfnamefont {A.}~\bibnamefont
  {Fleurence}}, \bibinfo {author} {\bibfnamefont {R.}~\bibnamefont
  {Friedlein}}, \bibinfo {author} {\bibfnamefont {T.}~\bibnamefont {Ozaki}},
  \bibinfo {author} {\bibfnamefont {H.}~\bibnamefont {Kawai}}, \bibinfo
  {author} {\bibfnamefont {Y.}~\bibnamefont {Wang}}, \ and\ \bibinfo {author}
  {\bibfnamefont {Y.}~\bibnamefont {Yamada-Takamura}},\ }\href {\doibase
  10.1103/PhysRevLett.108.245501} {\bibfield  {journal} {\bibinfo  {journal}
  {Phys. Rev. Lett.}\ }\textbf {\bibinfo {volume} {108}},\ \bibinfo {pages}
  {245501} (\bibinfo {year} {2012})}\BibitemShut {NoStop}%
\bibitem [{\citenamefont {Lin}\ \emph {et~al.}(2012)\citenamefont {Lin},
  \citenamefont {Arafune}, \citenamefont {Kawahara}, \citenamefont {Tsukahara},
  \citenamefont {Minamitani}, \citenamefont {Kim}, \citenamefont {Takagi},\
  and\ \citenamefont {Kawai}}]{Lin:2012zr}%
  \BibitemOpen
  \bibfield  {author} {\bibinfo {author} {\bibfnamefont {C.-L.}\ \bibnamefont
  {Lin}}, \bibinfo {author} {\bibfnamefont {R.}~\bibnamefont {Arafune}},
  \bibinfo {author} {\bibfnamefont {K.}~\bibnamefont {Kawahara}}, \bibinfo
  {author} {\bibfnamefont {N.}~\bibnamefont {Tsukahara}}, \bibinfo {author}
  {\bibfnamefont {E.}~\bibnamefont {Minamitani}}, \bibinfo {author}
  {\bibfnamefont {Y.}~\bibnamefont {Kim}}, \bibinfo {author} {\bibfnamefont
  {N.}~\bibnamefont {Takagi}}, \ and\ \bibinfo {author} {\bibfnamefont
  {M.}~\bibnamefont {Kawai}},\ }\href
  {http://stacks.iop.org/1882-0786/5/i=4/a=045802} {\bibfield  {journal}
  {\bibinfo  {journal} {Appl. Phys. Express}\ }\textbf {\bibinfo {volume}
  {5}},\ \bibinfo {pages} {045802} (\bibinfo {year} {2012})}\BibitemShut
  {NoStop}%
\bibitem [{\citenamefont {Quimbay}\ and\ \citenamefont
  {Strange}(2013{\natexlab{b}})}]{Quimbay:2013ab}%
  \BibitemOpen
  \bibfield  {author} {\bibinfo {author} {\bibfnamefont {C.}~\bibnamefont
  {Quimbay}}\ and\ \bibinfo {author} {\bibfnamefont {P.}~\bibnamefont
  {Strange}},\ }\href@noop {} {\  (\bibinfo {year} {2013}{\natexlab{b}})},\
  \Eprint {http://arxiv.org/abs/1311.2021} {arXiv:1311.2021 [quant-ph]}
  \BibitemShut {NoStop}%
\bibitem [{\citenamefont {Fl{\"u}gge}(1974)}]{Flugge:1974aa}%
  \BibitemOpen
  \bibfield  {author} {\bibinfo {author} {\bibfnamefont {S.}~\bibnamefont
  {Fl{\"u}gge}},\ }\href@noop {} {\emph {\bibinfo {title} {Practical quantum
  mechanics. {I}, {II}}}}\ (\bibinfo  {publisher} {Springer-Verlag},\ \bibinfo
  {address} {Berlin},\ \bibinfo {year} {1974})\BibitemShut {NoStop}%
\bibitem [{\citenamefont {Abramowitz}\ and\ \citenamefont
  {Stegun}(1964)}]{Abramowitz:1964aa}%
  \BibitemOpen
  \bibfield  {author} {\bibinfo {author} {\bibfnamefont {M.}~\bibnamefont
  {Abramowitz}}\ and\ \bibinfo {author} {\bibfnamefont {I.~A.}\ \bibnamefont
  {Stegun}},\ }\href@noop {} {\emph {\bibinfo {title} {Handbook of Mathematical
  Functions with Formulas, Graphs, and Mathematical Tables}}},\ \bibinfo
  {edition} {{ninth Dover printing, tenth printing of U.S. Governement Printing
  Office}}\ ed.\ (\bibinfo  {publisher} {Dover},\ \bibinfo {address} {New
  York},\ \bibinfo {year} {1964})\BibitemShut {NoStop}%
\bibitem [{\citenamefont {Chen}\ \emph {et~al.}(2012)\citenamefont {Chen},
  \citenamefont {Liu}, \citenamefont {Feng}, \citenamefont {He}, \citenamefont
  {Cheng}, \citenamefont {Ding}, \citenamefont {Meng}, \citenamefont {Yao},\
  and\ \citenamefont {Wu}}]{Chen:2012aa}%
  \BibitemOpen
  \bibfield  {author} {\bibinfo {author} {\bibfnamefont {L.}~\bibnamefont
  {Chen}}, \bibinfo {author} {\bibfnamefont {C.-C.}\ \bibnamefont {Liu}},
  \bibinfo {author} {\bibfnamefont {B.}~\bibnamefont {Feng}}, \bibinfo {author}
  {\bibfnamefont {X.}~\bibnamefont {He}}, \bibinfo {author} {\bibfnamefont
  {P.}~\bibnamefont {Cheng}}, \bibinfo {author} {\bibfnamefont
  {Z.}~\bibnamefont {Ding}}, \bibinfo {author} {\bibfnamefont {S.}~\bibnamefont
  {Meng}}, \bibinfo {author} {\bibfnamefont {Y.}~\bibnamefont {Yao}}, \ and\
  \bibinfo {author} {\bibfnamefont {K.}~\bibnamefont {Wu}},\ }\href {\doibase
  10.1103/PhysRevLett.109.056804} {\bibfield  {journal} {\bibinfo  {journal}
  {Phys. Rev. Lett.}\ }\textbf {\bibinfo {volume} {109}},\ \bibinfo {pages}
  {056804} (\bibinfo {year} {2012})}\BibitemShut {NoStop}%
\bibitem [{\citenamefont {Ezawa}\ and\ \citenamefont
  {Nagaosa}(2013)}]{Ezawa:2013aa}%
  \BibitemOpen
  \bibfield  {author} {\bibinfo {author} {\bibfnamefont {M.}~\bibnamefont
  {Ezawa}}\ and\ \bibinfo {author} {\bibfnamefont {N.}~\bibnamefont
  {Nagaosa}},\ }\href {\doibase 10.1103/PhysRevB.88.121401} {\bibfield
  {journal} {\bibinfo  {journal} {Phys. Rev. B}\ }\textbf {\bibinfo {volume}
  {88}},\ \bibinfo {pages} {121401} (\bibinfo {year} {2013})}\BibitemShut
  {NoStop}%
\bibitem [{\citenamefont {Ezawa}(2012)}]{Ezawa:2012aa}%
  \BibitemOpen
  \bibfield  {author} {\bibinfo {author} {\bibfnamefont {M.}~\bibnamefont
  {Ezawa}},\ }\href {\doibase 10.1103/PhysRevLett.109.055502} {\bibfield
  {journal} {\bibinfo  {journal} {Phys. Rev. Lett.}\ }\textbf {\bibinfo
  {volume} {109}},\ \bibinfo {pages} {055502} (\bibinfo {year}
  {2012})}\BibitemShut {NoStop}%
\bibitem [{\citenamefont {Ezawa}(2013)}]{Ezawa:2013ab}%
  \BibitemOpen
  \bibfield  {author} {\bibinfo {author} {\bibfnamefont {M.}~\bibnamefont
  {Ezawa}},\ }\href {\doibase 10.1103/PhysRevLett.110.026603} {\bibfield
  {journal} {\bibinfo  {journal} {Phys. Rev. Lett.}\ }\textbf {\bibinfo
  {volume} {110}},\ \bibinfo {pages} {026603} (\bibinfo {year}
  {2013})}\BibitemShut {NoStop}%
\end{thebibliography}%

\end{document}